\documentclass[10pt,twocolumn]{IEEEtran}
\usepackage{amssymb}
\usepackage{amsmath}
\usepackage[version=3]{mhchem}
\usepackage{graphicx}
\usepackage{multirow}
\usepackage{wrapfig}
\usepackage{color}
\usepackage{colortbl}
\usepackage{soul}
\usepackage{subfigure}
\usepackage{enumerate}
\usepackage{paralist}
\usepackage[normalem]{ulem}

\begin{document}
\title{On the Physical Design of \\Molecular Communication Receiver \\Based on Nanoscale Biosensors}
\author{Murat Kuscu,~\IEEEmembership{Student Member,~IEEE}
        and Ozgur B. Akan,~\IEEEmembership{Fellow,~IEEE}
        \thanks{The authors are with the Next-generation and Wireless Communications Laboratory (NWCL), Department of Electrical and Electronics Engineering, Koc University, Istanbul, 34450, Turkey (e-mail: \{mkuscu, akan\}@ku.edu.tr).}
        \thanks{This work was supported in part by the European Research Council (ERC) under grant ERC-2013-CoG \#616922.}}

\maketitle

\begin{abstract}
Molecular communications (MC), where molecules are used to encode, transmit, and receive information, is a promising means of enabling the coordination of nanoscale devices. The paradigm has been extensively studied from various aspects, including channel modeling and noise analysis. Comparatively little attention has been given to the physical design of molecular receiver and transmitter, envisioning biological synthetic cells with intrinsic molecular reception and transmission capabilities as the future nanomachines. However, this assumption leads to a discrepancy between the envisaged applications requiring complex communication interfaces and protocols, and the very limited computational capacities of the envisioned biological nanomachines. In this paper, we examine the feasibility of designing a molecular receiver, in a physical domain other than synthetic biology, meeting the basic requirements of nanonetwork applications. We first review the state-of-the-art biosensing approaches to determine whether they can inspire a receiver design. We reveal that nanoscale field effect transistor based electrical biosensor technology (bioFET) is a particularly useful starting point for designing a molecular receiver. Focusing on bioFET-based molecular receivers with a conceptual approach, we provide a guideline elaborating on their operation principles, performance metrics and design parameters. We then provide a simple model for signal flow in silicon nanowire (SiNW) FET-based molecular receiver. Lastly, we discuss the practical challenges of implementing the receiver and present the future research avenues from a communication theoretical perspective.
\end{abstract}


\IEEEpeerreviewmaketitle

\section{Introduction}
\label{Introduction}
\IEEEPARstart{D}{esigning} fully functional nanoscale devices with sensing, computing and actuating capabilities has been a long-standing goal of science and engineering community. Recent advances in nanotechnology including the inventions of novel nanomaterials like carbon nanotube and graphene have enabled the miniaturization of various functional devices like computing and memories into nanoscale dimensions. This exciting progress has led to a thorough investigation on the feasibility of enabling communication between nanomachines, forming \textit{nanonetworks}, to realize more complex tasks for ground breaking applications such as collaborative in-body drug delivery and continuous health monitoring \cite{Akyildiz2008} \cite{Kuscu2015-2}.

Since implementing conventional electromagnetic communications among nanomachines are obstructed by the limitations of antenna sizes, researches have started a quest for finding alternative communication methods. Among the several paradigms proposed for use in nanonetworks, MC is the most promising one because it already exists in nature as the main communication mechanism of living cells and other microorganisms, and thus, its feasibility in nanoscale domain is readily proven. MC uses molecules to encode, transmit and receive information. One of the physical characteristics of molecules, such as type or concentration, is modulated to be transferred either through active propagation or diffusion-based passive propagation in a fluid medium \cite{Nakano2012} \cite{Kuran2011}.

A substantial body of work provides signal propagation and noise models for both active and passive propagation with various modulation schemes \cite{Akyildiz2013}. Many aspects of MC like addressing and routing have been widely investigated; and advanced communication protocols for transport, network and link layers have just begun to be designed \cite{Nakano2014}. However, so far, little attention has been given to the physical design of transmitter and receiver devices compatible with MC.

A limited number of works have focused on the signal processing aspects of molecular receiver with the aim of developing optimal detection schemes generally being adapted from the conventional communication theoretical tools \cite{Kilinc2013}. However, these studies ignore the design of the receiver antenna and the characteristics of the antenna's output signal which is to be processed. They implicitly assume that the output signal is already suitable, in terms of physical form, amplitude and resolution, for the proposed detection schemes, all of which require fast and complex signal processing with high computational resources. On the other hand, the current predictions on the design of nanomachines with MC ability are based on the synthetic biology approaches, and a widespread consensus is established on this completely bio-inspired vision \cite{Nakano2012} \cite{Nakano2014}. However, synthetic biology is currently far away from taking the full control over the functionality of a living cell or designing artificial cells that are capable of operating in a nanonetwork application. Moreover, relying only on synthetic cells limits the application range of MC to \textit{in vivo} operations. It is clear that novel approaches are required to remove the discrepancy between the envisioned applications and the device architectures. At this point, this paper aims to pave the way for the physical design of a nanoscale MC receiver compatible with the proposed communication schemes requiring high computational capacity.

An MC receiver should be capable of \textit{in-situ}, continuous, label-free and selective detection of molecular messages that arrive in the close vicinity of the nanomachine on which it is implemented. It can be considered to consist of a biorecognition unit selectively recognizing the information molecules; a transducer unit converting recognition events to a processable signal; and a processor unit analyzing the transducer output to extract the transmitted information based on a preset scheme. Functionality of an MC receiver is very similar to the one of biosensors, which are also designed for the aim of quantifying analyte concentrations in a solution \cite{Schellera2001}. Therefore, in this paper, we first review the state-of-the-art biosensing approaches to determine whether and in what extent they satisfy the design requirements of an MC receiver.

A careful examination of biosensing options, ranging from optical to mechanical and electrical methods, reveals that field-effect transistor-based electrical biosensors (bioFETs) are quite promising for the design of molecular receiver \cite{Poghossian2014}. Hence, in the second part of the paper, we focus on a conceptual FET-based molecular receiver. We provide a comprehensive design guideline comprising of the operation principles and the performance metrics of the device. We elaborate on the design options of recognition and transducer units for the optimization of the overall performance from the communication theoretical perspective. Moreover, we present a simple model for a SiNW FET-based MC receiver, and evaluate the receiver performance in terms of sensitivity and Signal-to-Noise Ratio (SNR). We examine integrability of the proposed receiver design into more advanced communication schemes including molecular division multiple access (MDMA), molecular shift keying (MoSK), spatial diversity combining. Lastly, we discuss the challenges before the implementation of the device from a practical aspect, and state the future research directions.

\section{Molecular Communications Receiver}
\label{MCreceiver}
\begin{figure}[!t]
\centering
\includegraphics[width=7.5cm]{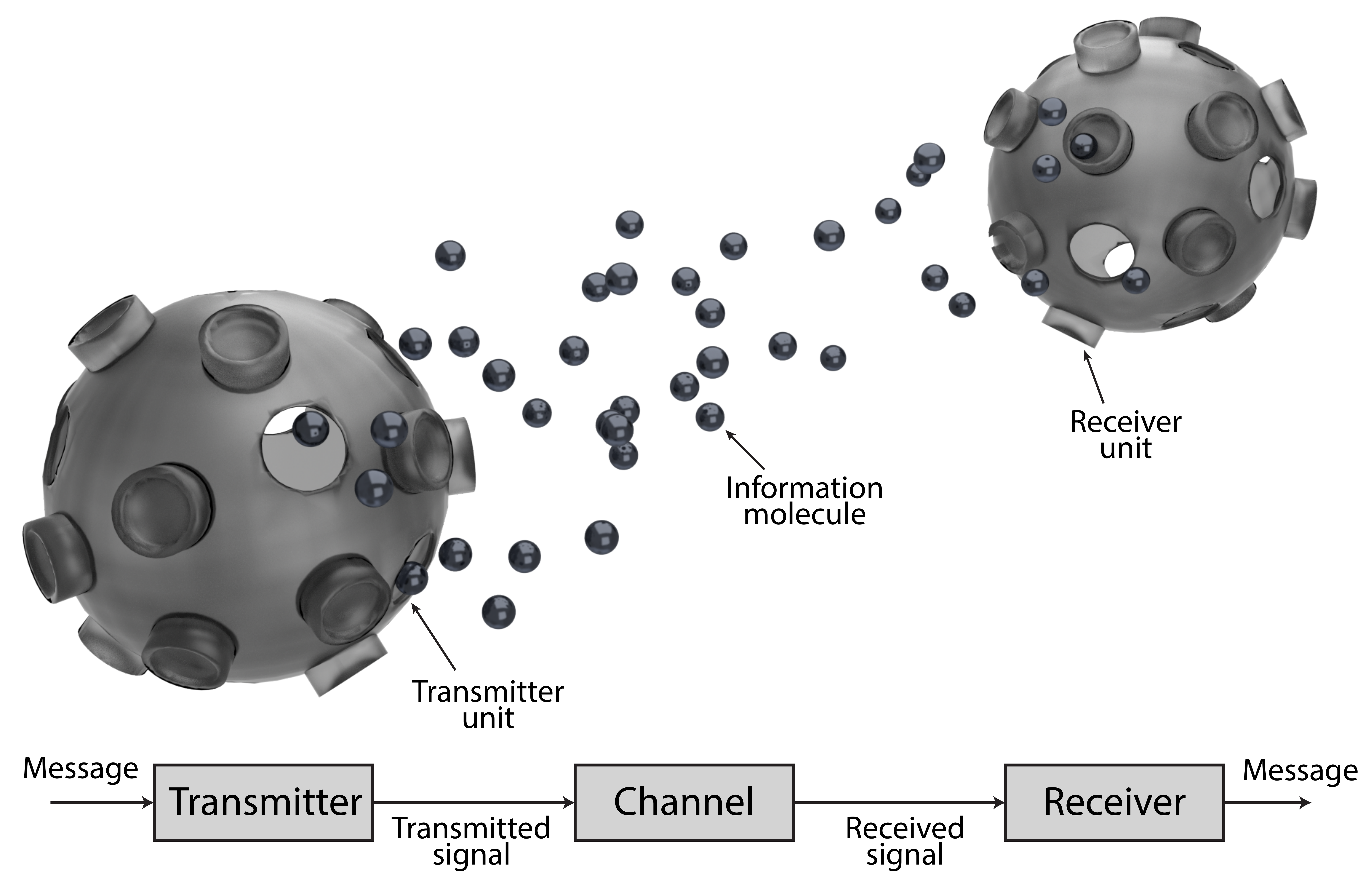}
\caption{MC between a pair of nanomachines.}
\label{fig:MolCom}
\end{figure}

MC between a single pair of nanomachines can be depicted as in Fig.~\ref{fig:MolCom}, where the molecular messages encoded into concentration propagate through diffusion from the transmitting nanomachine to the receiving nanomachine. Receiver of a nanomachine is responsible for detecting the incoming molecular messages, converting them into a processable signal, and processing the signal for extracting the encoded information. Information may then be used by the nanomachine to realize a prespecified operation. Therefore, the performance of the receiver is critical for the proper functioning of the nanomachine, and thus of the nanonetwork application.

\begin{figure}[!t]
\centering
\includegraphics[width=7.5cm]{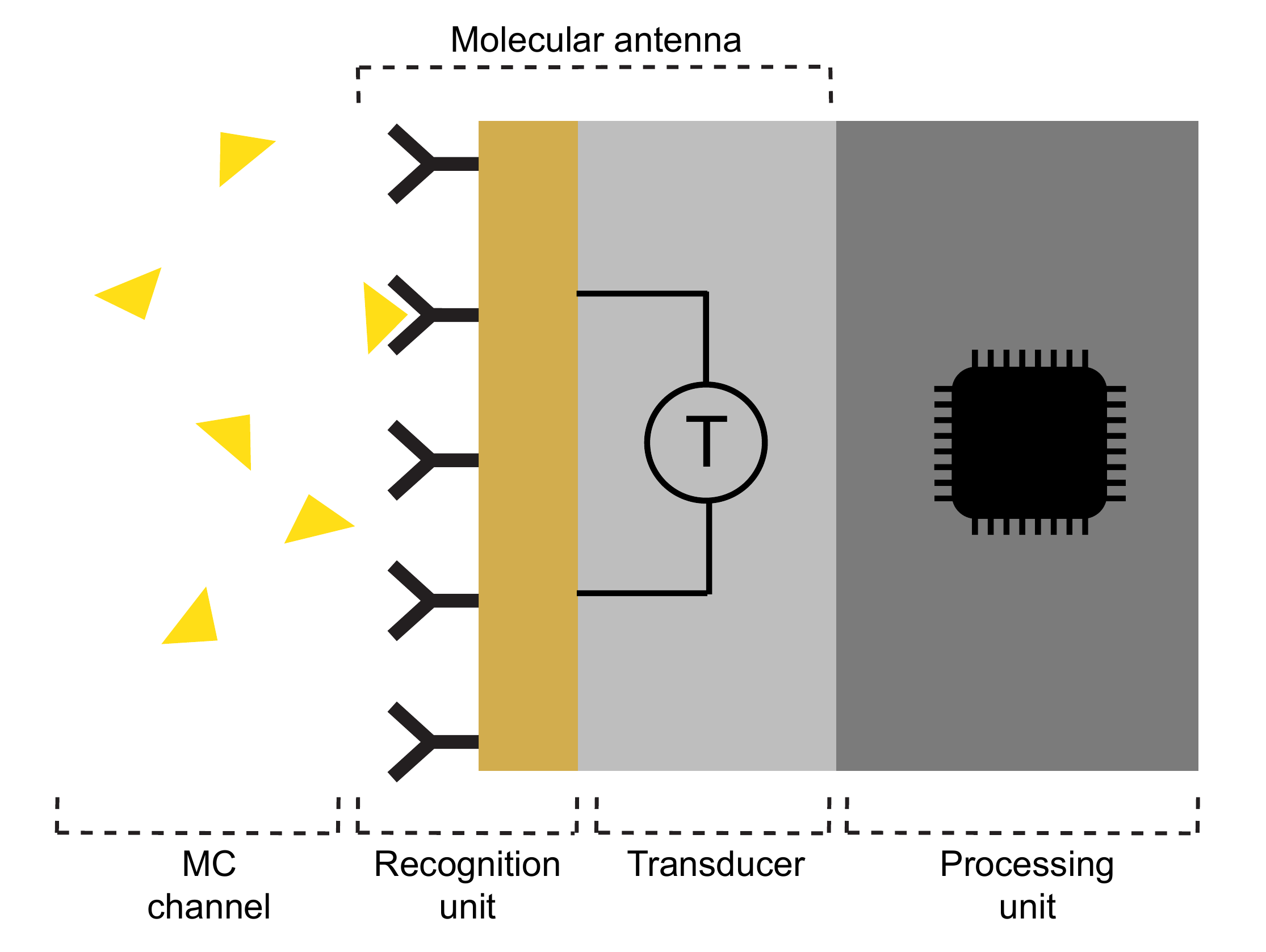}
\caption{Functional units of an MC receiver.}
\label{fig:Receiver}
\end{figure}

MC receiver slightly differs from an EM communication receiver. Fig.~\ref{fig:Receiver} demonstrates the block diagram of its envisioned architecture. It basically consists of two subunits, namely, a molecular antenna, and a processing unit which is coupled to the output of the antenna. Molecular antenna is comprised of a recognition unit and a transducer. Recognition unit is the interface between the communication channel and the receiver. Its function is to establish the selectivity only to the molecules that carry information, and thus to minimize the interference from other molecules in the communication environment. Transducer works intimately with the recognition unit, and converts the molecular recognition events into a processable form by means of electrical, optical or chemical signals. Transduced signals, i.e., output signals of the antenna, are simultaneously collected by the processing unit which then may amplify and demodulate the signal to recover the transmitted information based on a preset modulation scheme.

To be operable in an MC application, the receiver should basically satisfy the following requirements:

\begin{itemize}
\item \textbf{\textit{In situ} operation and in-device processing:}
Nanomachines are required to operate in a physically independent manner. If nanomachines are controlled through an external agent, the interface between them should not interrupt the system operation. In the same manner, an MC receiver integrated into a nanomachine should be capable of \textit{in situ} operation. Receiver should detect molecular messages through in-device processing, i.e., without requiring a post-processing of the transduced signals by an external macroscale device or a human controller.

\item \textbf{Label-free detection:}  
Receiver should be able to recognize the information carrying molecules based on their intrinsic characteristics, i.e., without requiring a molecular labelling procedure or any other preparation stage.

\item \textbf{Continuous operation:}
Detection should be continuous. This requires the recognition and transduction to be reversibly responsive, i.e., they should return to the initial state after detection to accept succeeding molecular messages without causing a communication error.

\item \textbf{Nanoscale dimensions:} The components of the receiver should have nanoscale dimensions for the integration of the overall receiver into a nanomachine.

\end{itemize}

In addition to these requirements, for \textit{in vivo} applications, the receiver should also be biocompatible. In other words, the operation of the receiver should not have any detrimental effect on the biological environment, or its performance should not be degraded by the environmental conditions.

\section{Overview of Biosensing Mechanisms}
\label{Overview}
Developing a biosensor capable of detecting the existence of a target molecule in an environment or quantifying the concentration of a molecular specie has been a long-standing goal of analytical chemistry. For this aim, many different biosensing approaches have been proposed for a plethora of target molecular species ranging from proteins to oligonucleotides, as well as small molecules like glucose \cite{Perumal2014}.

A biosensor, in general, consists of a biorecognition layer and a transducer which converts the recognition events into a processable form. It is not surprising that a biosensor and a molecular receiver have common design principles, since they both aim to analytically recognize the molecules. However, there are also fundamental differences between them. First, it is not sufficient for an MC receiver just to detect the presence of a molecule in an equilibrium state as for the case of a biosensor. The receiver should be capable of detecting the information, which is encoded into a physical property related to the molecules, such as concentration, type, or arrival time, by continuously observing the environment. This requires communication theoretical tools to be applied. Another critical difference arises from the application domain. The biosensor literature is mostly directed to the design of biosensing approaches suitable for laboratory applications that necessitate the operation of macroscale readout devices and human observers due to the lack of an integrated processor. Moreover, biosensors have not yet been completely scaled down to nanoscale dimensions to be integrated into an envisioned nanomachine. Nevertheless, MC receiver design can be inspired from a biosensing method.

Biosensors proposed so far, in general, can be classified into three categories based on the transduction mechanism: electrical, optical, and mechanical biosensors. Electrical biosensing is based on the alterations in the current, voltage, or conductance in the transducer resulting from a chemical reaction or a binding event realized in the recognition unit \cite{Grieshaber2008}. Similarly, optical sensing exploits the changes in one of the optical characteristics of the transducer upon the recognition. Optical transduction and recognition are generally realized on the same unit. Several optical sensing mechanisms are developed based on surface plasmon resonance (SPR), chemiluminescence, fluorescence and FRET \cite{Borisov2008}. On the other hand, mechanical sensing benefits from modulation of mechanical resonance with binding of molecules to the transducer surface \cite{Tamayo2013}.

In addition to these engineered sensing methods, a fourth mechanism which is called biochemical or molecular transduction is prevalent in nature as part of the cellular communications \cite{Cantrell1996}. The mechanism converts the biorecognition events on the cellular membrane into an ion flux through the membrane pores of the cell to trigger a function with the potential change, or the events activate the secondary messengers in the cytoplasm to realize cellular functions.

Considering the basic requirements of the MC receiver, we suggest focusing on electrical biosensing approaches. The reason for eliminating the optical and mechanical biosensing options is that they rely on macroscale detectors that cannot be accepted for an MC receiver considering the requirement of \textit{in situ} operation. Although the biorecognition and transduction units can be downsized to nanoscale, these units should be excited by optical lasers, and the resultant signals, for both optical and mechanical sensing, should be observed through optical detectors to extract the sensing information. Even for bioluminescence-based methods which do not require excitation of the transducer, there is currently no approach towards realizing \textit{in situ} processing of the resultant optical signals mainly due to the diffraction limit of light.

To compare the suitability of electrical and biochemical sensing for an MC receiver, we also need to consider the corresponding nanomachine architectures, which are in the nanobioelectronic and synthetic-cell domain, respectively. In this context, the advantages of electrical sensing are several. First, there is a clear trend towards the miniaturization of electrical processing units into nanoscale. Since electrical biosensors intrinsically provide processable inputs for these devices, they can be readily integrated in nanobioelectronic domain. On the other hand, the assumption of synthetic cell-based nanomachines, which is based on the combination of the biochemical transduction and high processing capabilities, seems very vague. Biochemical sensing implies also biochemical processing. Electrical operations is incomparably faster than biochemical processes, thus, we can assert that electrical sensing also promises for fast processing of the transduced signals. Moreover, electrical sensing can provide the receiver with much more diverse functionality which can be tuned and controlled electrically in a similar way with the conventional counterparts. Also, interfacing with macroscale equipments in a wireless manner is more feasible for the electrical domain. Although synthetic-cell based nanomachines intrinsically provide ultimate biocompatibility; with the advances in bioelectronics, it is not a difficult problem also for the nanobioelectronic devices. Therefore, comparing the state-of-the-art in the two research domains, obviously electrical nanomachines seem more suitable and feasible for the envisaged nanonetwork applications. This leads us to investigate, in more details, the electrical sensing approaches to be adapted for an MC receiver.

Electrical biosensors can be grouped into two classes according to their recognition method: biocatalytic and affinity-based sensors. Biocatalytic recognition employs specific enzymes immobilized on a layer to detect the targets based on enzymatic reactions \cite{Wang2008}. When a target is bound to an enzyme, an electroactive specie, e.g., hydrogen ion, is generated as a result of the reaction. If the specie achieves to diffuse into the working electrode of the transducer, it modulates one of the electrical characteristics of the device. Commercialized macroscale glucose sensors can be given as example to the biocatalytic electrical sensors. On the other hand, affinity-based sensing is based on the selective binding of certain receptor-ligand pairs on the recognition layer \cite{Rogers2000}.

Biocatalytic techniques require two-step reaction, i.e., enzymatic reaction and consecutive detection of the produced electroactive species. This complicates the procedure as the device gets smaller. Also, generation of additional products may not be desired during the operation. Additionally, the method is limited to specific analytes that generate the electroactive species upon the reaction with the corresponding enzymes. On the other hand, affinity-based sensing uses the nature's approach for recognition offering a more direct detection scheme. Also the method has proven feasible for a larger range of target molecules with higher sensitivity and selectivity. Moreover, affinity-based biorecognition is more convenient for the current molecular communication models which are mostly based on ligand-receptor binding, thus, provides an easier path to be adapted into the receiver models. Therefore, we focus our attention into affinity-based electrical sensing.
\begin{figure}[!t]
\centering
\includegraphics[width=9cm]{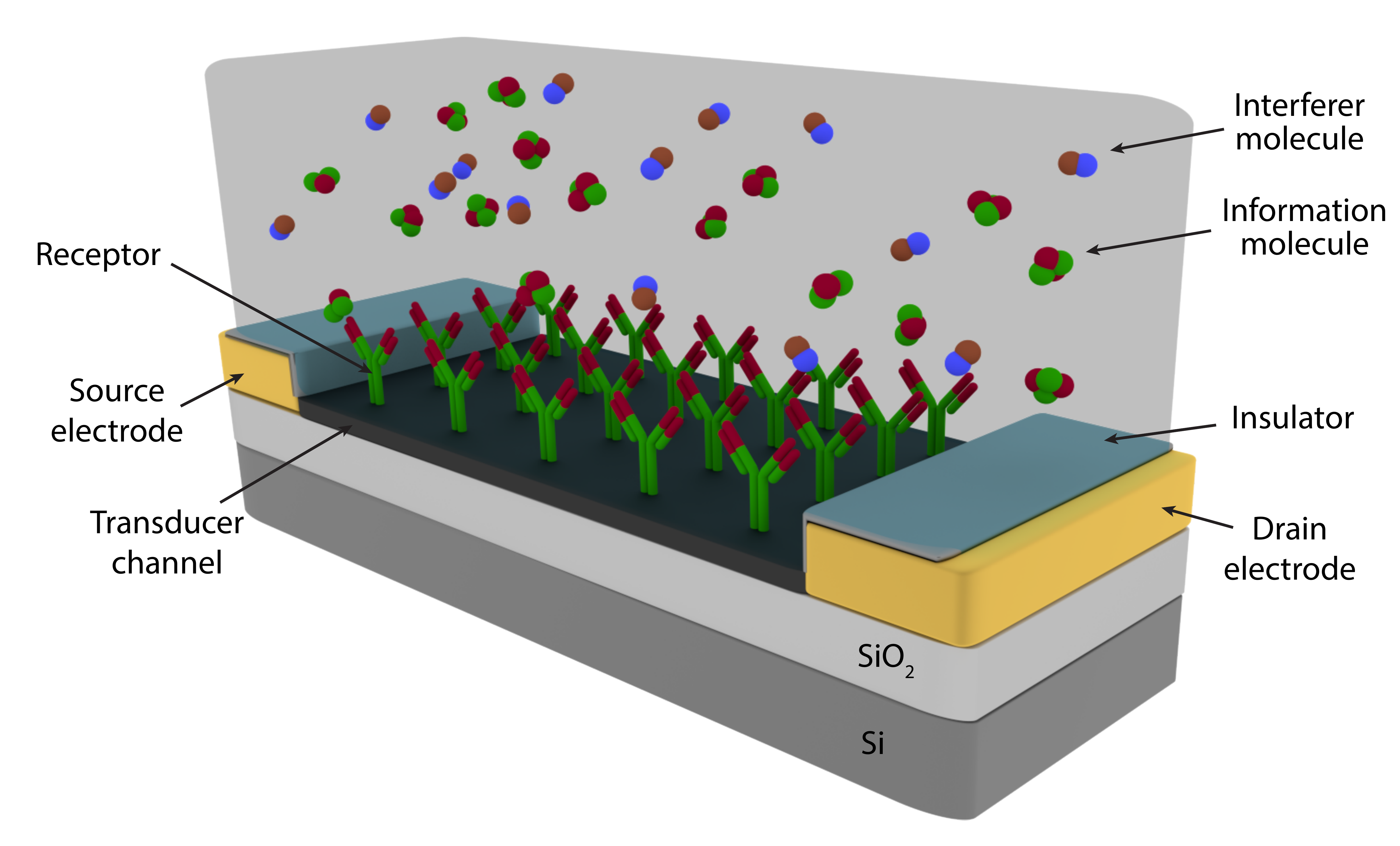}
\caption{Conceptual design of FET-based MC receiver.}
\label{fig:Design}
\end{figure}

Affinity-based sensing is also classified depending on the detection format in the transducer unit: amperometric, potentiometric, conductometric, and impedance-based \cite{Rogers2000}. Briefly, amperometric detection is based on the changes in the current; potentiometric detection relies on the change of the potential between two reference electrodes; conductometric detection is based on the modulation of the conductance by the recognition events; and impedance-based detection is a frequency-domain method benefitting from the impedance variations as the result of molecule binding events. These methods are developed primarily for \textit{ex situ} analysis in laboratory conditions; thus, they do not promise for portable and \textit{in situ} applications. Fortunately, advances in nanotechnology in the last decade have led to the design of a new method for affinity-based electrical sensing which is based on FET technology using nanowires, nanotubes, organic polymers, and graphene, as the transducer unit \cite{Yang2010} \cite{Curreli2008}. It is based on the modulation of transducer conductivity by the electrostatic effects resulting from the bindings on the recognition layer. FET-based biosensors (bioFETs) promises for label-free, continuous and \textit{in situ} operation in nanoscale dimensions, thus, the design of a bioFET stands as a good starting point for the design of an MC receiver. Hence, in the remaining of the paper, we elaborate on the FET-based molecular receiver.
\begin{figure}[!t]
\centering
\includegraphics[width=7.5cm]{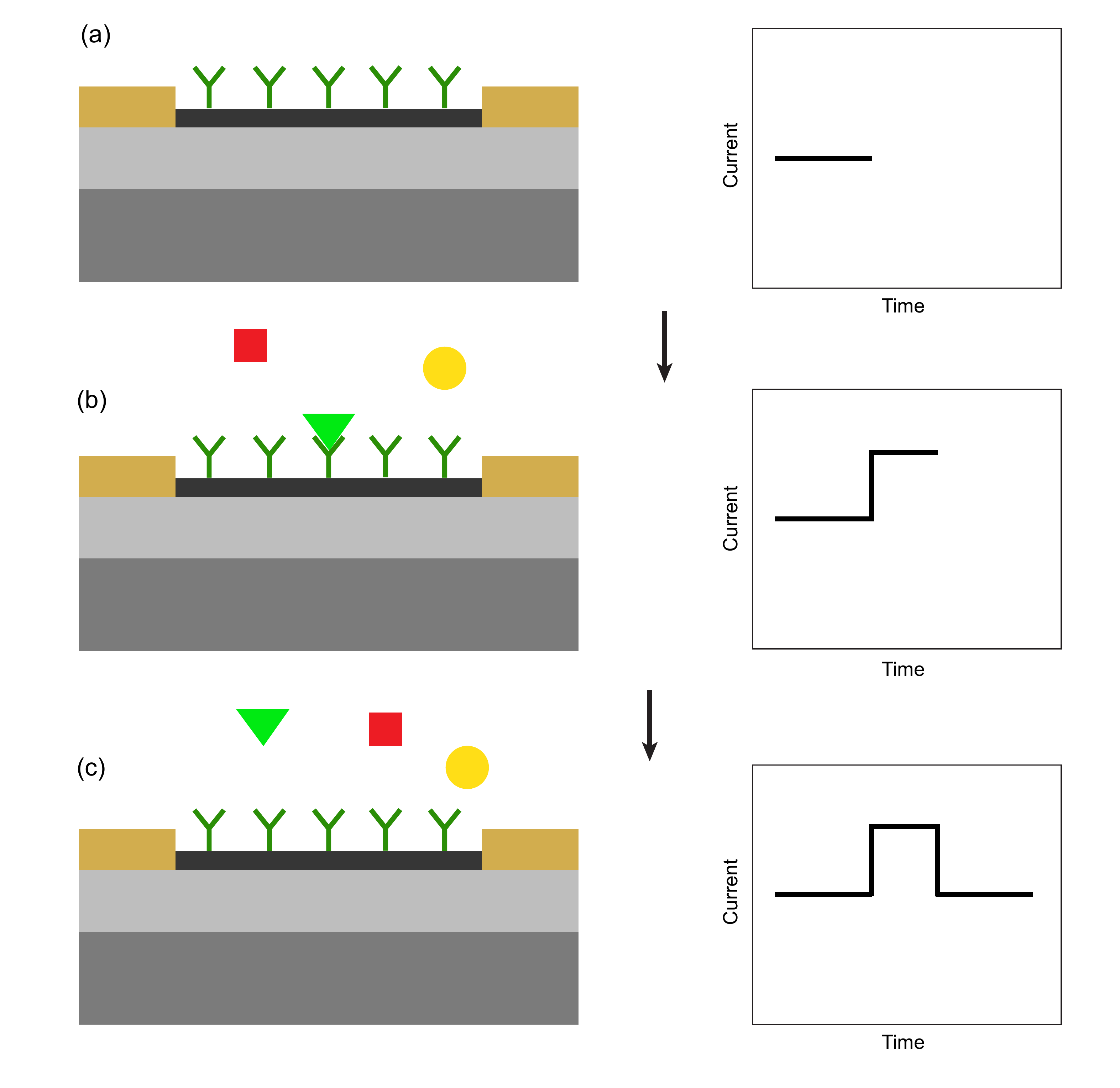}
\caption{Expected time domain response of a p-type doped FET-based MC receiver in terms of drain-to-source current in the transducer channel, when (a) no information molecule is present in the reception space, (b) a negatively charged information molecule binds to a surface receptor, (c) the bound molecule leaves the recognition unit.}
\label{fig:Response}
\end{figure}

\section{FET-Based Molecular Receiver}
\label{FETbased}
In conventional FET type transistors, current flows between source and drain electrodes through a semiconductor channel whose conductance is controlled via a perpendicular electric field created by a gate electrode. Conductivity is proportional to the carrier density accumulated in the channel by the electric field, and reflected to the changes in the voltage-current characteristics. Replacing the gate electrode with a biofunctionalized surface through the immobilization of affinity-based receptors directly on top of the channel as depicted in Fig.~\ref{fig:Design}, the device becomes a biosensor, namely bioFET.

Binding of charged analytes to the surface receptors results in accumulation or depletion of the carriers on the semiconductor channel, and modulates the channel conductivity; thus, the conductivity becomes a function of the analyte density and the amount of analyte charges. For example, if the channel is p-type doped, negatively charged analytes accumulate the positive hole carriers in the channel when they bind to the receptors; as a result, the conductivity increases depending on the density of the bound analytes as shown in Fig.~\ref{fig:Response}. The conductivity change is reflected to the current ($I_{DS}$) flowing through the channel if a constant source-to-drain potential ($V_{DS}$) is maintained. For applications in ionic solutions, usually, a reference electrode is immersed in the solution to stabilize the surface potential of the channel on its base level which may be highly fluctuated due to undesirable ionic adsorptions complicating the detection of target analytes \cite{Nair2007}. It is clear that FET-based biosensing enables direct, label-free, continuous and \textit{in situ} sensing of the molecules by not requiring any complicated processes like labelling of the molecules or the use of any macroscale equipment for readout and processing operations.

Simple operation principles together with the extensive literature on FETs established through many years, electrical controllability of the main device parameters, high-level integrability, and plethora of optimization options for varying applications make FET-based biosensing technology also a quite promising approach for electrical MC receiver. Therefore, we direct our attention to devise a molecular receiver based on the principles of affinity-based bioFETs.
\subsection{Signal Flow and Noise Sources}
\label{Principles}
To characterize the operation of a FET-based molecular receiver, we need to investigate the input-output signal relations together with the expected contributions of the noise sources, which result in random fluctuations in the electrical output signal. Noise sources for a FET-based MC receiver can be divided into three categories \cite{Hassibi2007}:
\subsubsection{Reception Noise}
A molecule in the reception space undergoes random motion which is governed by Brownian dynamics, and stochastically binds to the receptors on the recognition layer. The uncertainty in the location and the binding state of the molecules results in random fluctuations of the transduced signal and may severely hamper the instant detection of the concentration, and thus, the molecular messages \cite{Hassibi2005}. This type of reception noise is well-studied from the MC perspective \cite{Pierobon2011}.

\subsubsection{Transducing noise}
Transducing operation also includes contributions from noise sources which can be classified into two types: thermal noise and $1/f$ (flicker) noise \cite{Rajan2013}, \cite{Deen2006}. Thermal noise is resulting from thermal fluctuations of charge carriers on the bound ligands. On the other hand, $1/f$ noise is caused by the traps and defects in the semiconductor channel, and could be effective at low frequencies where MC receiver is expected to operate. Hence, $1/f$ noise may dominate over other noise sources, and needs a careful investigation.

\subsubsection{Biological Interference}
For crowded environments, it is possible that molecules of other species, which show similar affinities for the receptors, exist in the reception space, thus, occasionally bind to the receptors and complicate the detection. This biological interference, which is also termed the background noise, could severely deteriorate the detection performance as the concentration and the charge of the interferers are comparably high \cite{Hassibi2005}. Biological interference should not be confused with the intersymbol interference (ISI) or co-channel interference which have been already investigated in \cite{Pierobon2012}, \cite{Pierobon2014}, from the aspect of MC. Note that the randomness of interferers' motion and binding to the receptors also contributes to the total reception noise of the receiver.

\begin{figure}[!t]
\centering
\includegraphics[width=9cm]{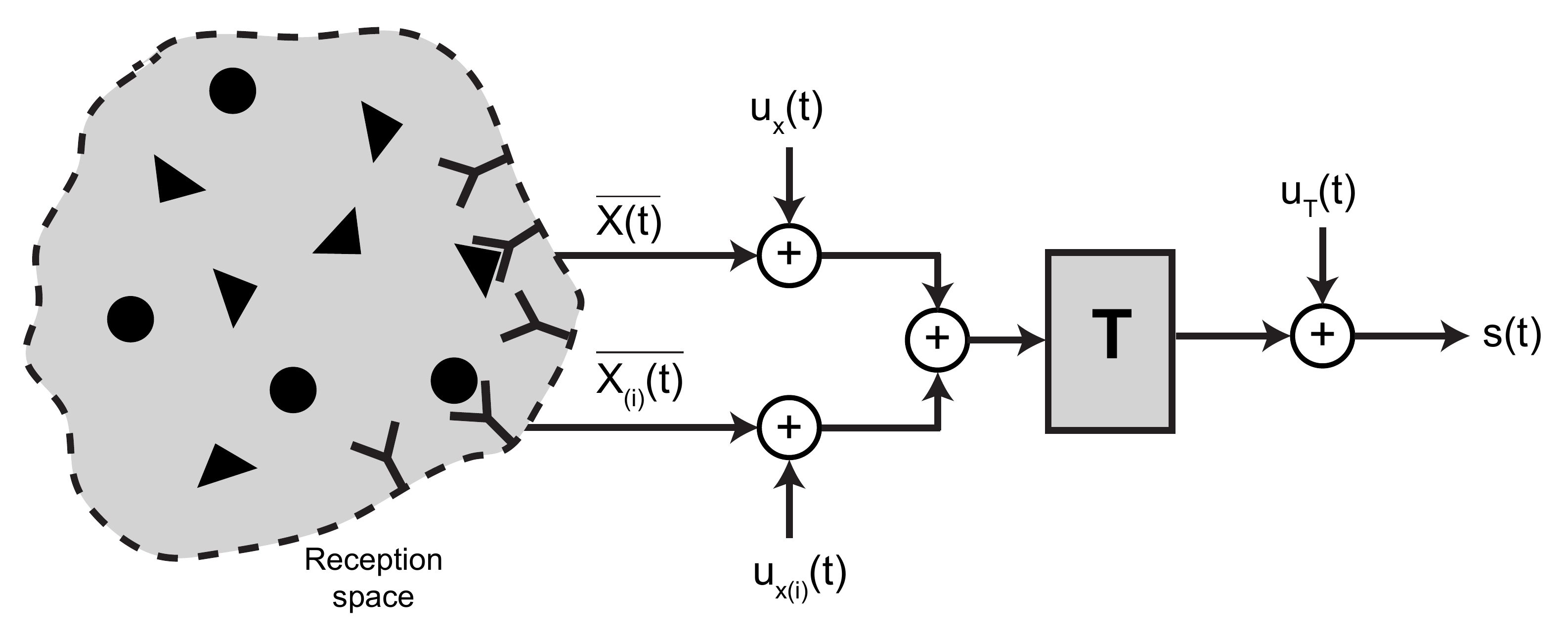}
\caption{Signal flow diagram of an MC receiver antenna. The triangles and circles denote the information and interferer molecules, respectively.}
\label{fig:SignalFlow}
\end{figure}

Incorporating the contribution of noise sources, a simple signal flow diagram of the receiver can be given as in Fig.~\ref{fig:SignalFlow}, assuming that the binding of interferer molecules and messenger molecules are uncorrelated. In fact, this is a loose assumption; because the receptor-ligand binding kinetics is expected to be affected by the number of bound interferer molecules. However, when the concentration of the receptors in the recognition layer is high enough compared to the total concentration of ligand and interferer molecules in the reception space, this assumption holds true. Then, the electrical output signal can be expressed as follows:
\begin{equation}
s(t) = T \overline{X(t)} +  T u_x(t) + u_T(t) + \sum_i^m T_{i} \left( \overline{X_{(i)}(t)} + u_{x(i)} \right), \label{OutputSignal}
\end{equation}
where $X(t)$ is a stochastic state matrix which represents the locations of the ligands, including the conjugated ones, in the reception space \cite{Hassibi2007}. The matrix relates the molecular density of the reception space to the occupancy rate of the receptors on the recognition layer. $T$ is the transducing vector which operates on the state matrix, and $\overline{X(t)}$ is the ensemble average of the state matrix. The $u_x(t)$ and $u_T(t)$ denote the reception noise and the total transducing noise, respectively. $\overline{X_{(i)}(t)}$ is the ensemble average of the state matrix of the $i$th interferer molecule, and $u_{x(i)}$ is the corresponding reception noise. $T_{i}$ is the transducing vector for the $i$th interferer. Note that, the reception noise resulting from the stochastic binding events is amplified by the transducing operation.

\subsection{Performance Metrics}
\label{Performance}

\subsubsection{Sensitivity}
Sensitivity is the capability of the receiver to perceive the small differences in the molecular concentration \cite{Nair2007}. It is a critical performance metric, especially for the design of MC systems which encodes the information into molecular concentrations, e.g., MC with Concentration Shift Keying (CSK). It can be defined as the change in the transducer current per change in the molecular concentration in the reception space of the receiver for a concentration range where linear operation can be assumed. Linear operation corresponds to the operation range where the receptors on the receiver surface are not saturated by the information molecules.

\subsubsection{Selectivity} 
The capability of the receiver to uniquely detect the information molecules in the reception space, which is probably crowded with interferers, is the measure of its selectivity \cite{Nair2007}. It is clear that the recognition unit is directly in charge of the selectivity as it sets the affinity with messenger ligands and possible interferers. Higher selectivity implies that the probability of the interferer molecules to bind the receptors is lower than the one of the information carrying molecules. Selectivity is a critical design consideration especially for MoSK-based MC systems \cite{Kuran2011}.

\subsubsection{Signal-to-Noise Ratio (SNR)}
SNR at the antenna output determines how easy the receiver can recover the transmitted signal in the presence of noise sources. It is a crucial metric to evaluate the reliability of the detection. It can be simply calculated as the ratio of the received signal power to the total noise power, i.e., $SNR = P_{signal}/P_{noise}$, in a given operation bandwidth of the receiver.

\subsubsection{Limit of Detection}
The minimum molecular concentration in the reception space required for the receiver to detect the existence of an information molecule is termed the limit of detection (LoD) \cite{Rajan2013-2}. It is not to be confused with the minimum concentration required for the exact determination of the concentration. Once the limit is determined, it can be exploited by the transmitter to save on the number of molecules transmitted, especially when the information is encoded into the molecule type as in MoSK \cite{Kuran2011}. The limit of detection is mainly set by the noise level of the overall reception process and the flow conditions of the environment.

\subsubsection{Temporal Resolution}
Temporal resolution defines how fast the receiver is able to sample the molecular concentration in the reception space. Considering that electrical processes are very fast compared to molecular processes, the limiting factor for the temporal resolution is expected to be the diffusion and binding kinetics. In order for the receiver to be able to detect all the messages carried by molecules into reception space, biorecognition should be realized in transport-limited manner. In other words, the binding kinetics should not be a limiting factor on the sampling rate. We further discuss the effect of binding kinetics on the temporal resolution in Section \ref{Receptors}.

In addition to these metrics, there are some other measures that can be used to quantify the performance of the electrical MC receiver. For example, the amount of information about the received molecular signal contained in the electrical output of the receiver is a quite important metric to determine the detection performance. In the information theory, this is represented by the mutual information between the receiver output and the incoming molecular signals. Mutual information is determined based on the conditional entropy, i.e., equivocation, of the received signal given the receiver's electrical output. These information theoretical metrics would be especially important for the design of optimal detection algorithms. They are needed for determining the end-to-end communication capacity of an MC system and developing encoding schemes that can achieve this capacity. However, they require a probabilistic model of the overall reception process combined with the random propagation of the information signals in MC channel. Likewise, if the receiver is implemented along with a differential amplifier, Common Mode Rejection Ratio (CMRR), which measures the capability of the device to reject the common modes in the incoming signals, will become an important metric to determine the receiver performance.

\section{Design Options}
\label{Design}
The performance of the receiver should be optimized by the design parameters under different application scenarios. This section elaborates on the main parameters corresponding to the two main receiver components, i.e., the transducer and the biorecognition unit.

\subsection{Semiconductor Channel}
\label{Channel}
\begin{figure}[!t]
\centering
\includegraphics[width=6cm]{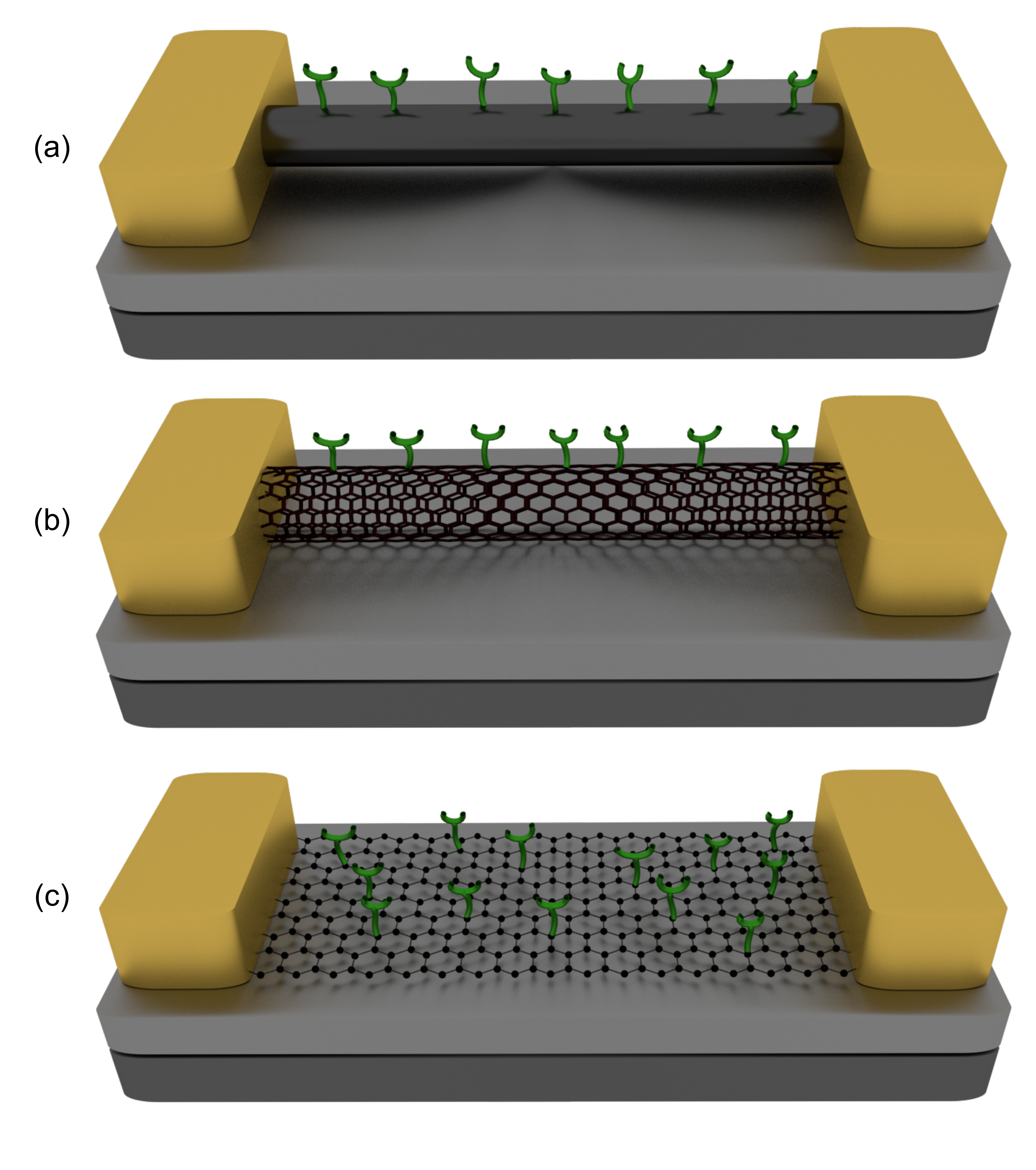}
\caption{FET-based MC receiver with different type of transducer channels: (a) SiNW channel, (b) SWCNT chanel, (c) graphene channel.}
\label{fig:Channel}
\end{figure}

Semiconductor channel between source and drain is the main transducing element of the receiver. Therefore, its electrical characteristics are critical for the performance of the receiver. $1/f$ noise is mainly resulting from the defects and traps on the channel. Moreover, the channel geometry is directly relevant to the surface coverage of the receiver, and affects its integrability to a nanomachine. Many nanomaterials have proven suitable for use in bioFET channel, such as silicon nanowires (SiNWs) \cite{Patolsky2005}, single walled carbon nanotubes (SWCNTs), graphene \cite{Yang2010}, molybdenum disulfide (MoS$_2$) \cite{Sarkar2014}, and organic materials like conducting polymers \cite{Torsi2013}. These materials, some of which are demonstrated in Fig.~\ref{fig:Channel}, are also candidates for the transducer channel of MC receiver.

First generation bioFETs rely on the use of one dimensional materials like SWCNT and SiNW as the channel in a bulk form. However, use in the form of single material or aligned arrays have proved to surpass the performance of the bulk channels in terms of sensitivity and reduced noise \cite{Curreli2008}. SWCNT-based bioFETs have attracted more attention due to the excellent electrical characteristics of carbon nanotubes which lead to higher sensitivity; however, clean fabrication of SWCNT without defects is the most challenging among all of the candidates, which may hamper its proper practice in an MC receiver \cite{Maehashi2009}. On the other hand, SiNWs can be fabricated in a relatively easy and controlled manner, with adjustable length, diameter, and doping levels which have direct influence on the material conductance \cite{Chena2011}. However, SiNWs still possess substantial production challenges due to their 1D nature.

Graphene has emerged as an alternative with its 2D planar structure providing higher spatial coverage in a single device with the possibility of immobilizing higher number of receptors \cite{Park2012}. The same 2D structure also leads to a higher sensitivity compared to 1D materials, since all of its atoms are able to closely interact with the bound molecules on the recognition layer. Moreover, intrinsic flexibility of graphene provides higher chance of integration into devices with non-planar surfaces which can be more suitable for the design of nanomachines in an MC application \cite{Yang2010}. Another 2D alternative for the transduction channel is MoS$_2$ which has been very recently shown to be more sensitive than the graphene channel \cite{Sarkar2014}. The possession of a bandgap which is not the case for graphene has been pointed out as the reason for the higher sensitivity of MoS$_2$-based bioFETs.

Organic conductive polymers also have been shown to provide similar electronic performances with the inorganic counterparts when used as the semiconducting channel. They can be utilized in both 1D and 2D forms \cite{Lin2012}. In addition, they provide ultimate biocompatibility which make them more preferable for \textit{in vivo} applications of MC.

The selection among these materials should be carefully made considering the requirements of the MC application \cite{Yang2010}. Also, the architecture of the overall nanomachine can mandate the use of flexible planar materials. On the other hand, some applications may require more complex receivers with more than one channels for multiplexing purposes which will be discussed in Section \ref{Advanced}; and for such cases, 1D materials may be more suitable allowing dense channel deployments.

\subsection{Bioreceptors and Ligands}
\label{Receptors}
Bioreceptor is another design fundamental which is the principal determinant of the receiver's selectivity. The array of receptors constitutes the interface between the receiver and the MC channel, being the active components of the recognition unit. The design considerations on the receptor molecules cannot be isolated from the information molecules, i.e., ligands; since the modulation of the transducer current is realized as a result of the binding of the ligands and the receptors to each other. The binding reaction can be simply represented as
$\ce{N_R + N_L  <=>[k_{+}][k_{-}] N_{B}}$,
where $N_R$, $N_L$, and $N_B$ denote the number of unbounded receptor, ligand, and receptor-ligand complexes, respectively; $k_+$ and $k_-$ are the association and dissociation reaction rates. The affinity between the ligands and the receptors is quantized by the dissociation constant which can be given by $K_D = k_{-}/k_{+}$. Affinity is inversely proportional to $K_D$. In order for the receiver to be selective only to the ligands that represent the information, the affinity of the receptors for the ligands should be significantly higher than their affinity for the interferer molecules that may exist in the reception space.

Mainly three types of recognition pairs used in affinity-based FET-type biosensors are feasible for an MC application: antibody/antigen (Ab/Ag), natural receptor/ligand, and aptamer/ligand \cite{Poghossian2014} \cite{Rogers2000}. These pairs are suitable because they bind to each other reversibly, and their sizes are small enough to be utilized in an MC application.

Antibodies are Y-shaped immune system proteins that are generated to bind to specific antigens which are perceived as the proteins of an harmful microorganism inside the body. Antibodies generally operate with lock-and-key principle, thus, provide high level selectivity for the corresponding antigens \cite{Curreli2008}. Antibodies are usually generated by introducing antigens into an animal. Therefore, their utilization may face production challenges and biocompatibility risks for \textit{in vivo} applications.

Natural receptors, i.e., neuroreceptors, taste and olfactory receptors, are cellular proteins immobilized within the plasma membrane. They have proven to be successfully implemented on the recognition unit of bioFETs to detect neurotransmitters, taste molecules, and odorants \cite{Song2014}. Since these receptors naturally operate in living organisms, their utilization in \textit{in vivo} applications do not bring up any biocompatibility problem. Moreover, their ability to detect odorants promises for long range MC applications reviewed in \cite{Gine2009}.

Additionally, aptamers, which are artificial single-stranded DNAs and RNAs, have also been widely utilized as recognition elements in bioFETs to detect a wide range of targets, i.e., ligands, including small molecules, proteins, ions, aminoacids, or other oligonucleotides. An aptamer for a specific ligand is selected through the SELEX process which is based on scanning a large library of DNAs and RNAs to determine a convenient nucleic acid sequence \cite{Luzi2003}. Availability of almost infinite number of different aptamer-ligand combinations with varying affinities makes aptamer-based recognition promise for highly selective MC receivers.

The characteristics of the ligand-receptor pairs affect the performance of the receiver, especially its sensitivity. The range of distance from the surface of the receiver where the transducer can detect a recognition event is limited because of the screening of the ligand charges by the ions existing in the communication environment \cite{Nair2008}. The mean effective charge of a free electron on a ligand as observed by the transducer is degraded as the distance between the ligand and the transducer increases. The relation is given by
\begin{equation}
q_{eff} = q e^{-\frac{r}{\lambda_D}}, \label{qeff}
\end{equation}
where $q$ is the elementary charge, and $r$ is the average distance of the ligand electrons to the transducer's surface \cite{Rajan2013}. $\lambda_D$ is the Debye length which quantizes the ionic strength of the solution according to the following relation
\begin{equation}
\lambda_D = \sqrt{\frac{\epsilon_R k_B T}{2 N_A q^2 c_{ion}}}, \label{eq:debye}
\end{equation}
where $\epsilon_R$ is the permittivity of the solvent, $k_B$ is the Boltzmann's constant, $T$ is the temperature, and $N_A$ is Avogadro's number, $c_{ion}$ is the ionic concentration of the medium \cite{Rajan2013}. Debye length is a key parameter in determining an appropriate receptor-ligand pair. The distance of the receptor part, where the charged ligand binds, from the transducer surface should not exceed the Debye length. Otherwise, charges of the bound ligands cannot be effectively reflected to transducer current. Therefore, receptors with lengths exceeding the Debye length should be avoided to develop a sensitive receiver. Aptamers and natural receptors with lengths typically smaller than 2nm are advantageous compared to antibodies \cite{Luzi2003}. On the other hand, antibodies can be shortened using only their fab fragments which are the main recognition elements \cite{Elnathan2012}.

Intrinsic charge of ligands also should be high enough for a proper receiver operation. Most biomolecules like proteins and nucleic acids are highly (and negatively) charged in physiological conditions, but the amount of charge may significantly alter depending on the pH of the environment \cite{Nair2007}.

Additionally, density of the receptors can remarkably affect the performance. It is obvious that dense deployment of the receptors on the receiver surface increases the chance of binding to a ligand. However, the receptor density should be optimized together with the concentration of ligands used for communication. On one hand, if the ligand concentration is very low compared to the receptor density, then, most of the receptors will be unbounded; and thus, prone to the binding of the interferers, which may result in higher background noise. On the other hand, if the information is represented by very high concentrations of ligands compared to the receptor density, then, the recognition unit may face saturation, and the linear operation range may be disturbed, which then complicates the detection.

\section{A Modeling Approach for SiNW FET-Based MC Receiver}
\label{Numerical}
In Section \ref{Principles}, we gave an expression in \eqref{OutputSignal} to describe the signal flow for the operation of a general MC receiver. That expression includes a transducer vector and a stochastic state matrix denoting the random locations and binding/unbinding states of information molecules. Although it gives an intuition for the receiver operation, it is only useful for simulation-based approaches, and it does not lead to the derivation of analytical expressions for deterministic signal flow to evaluate the performance of the receiver.

In this section, based on the existing modeling efforts in the literature \cite{Berezhkovskii2013} \cite{Kuscu2015c}, we present an analytical model of molecular signal reception for a SiNW FET-based MC receiver and evaluate its performance in terms of mean electrical response, sensitivity and SNR. For modulation of molecular signals, we assume that CSK is utilized such that the messages are represented by different levels of ligand concentration. The receiver detects the concentration-encoded messages by observing the transducer channel current.

\subsection{Model of Biorecognition Unit}
For the analytical model of biorecognition unit, following assumptions are made:
\begin{inparaenum}[\itshape i\upshape)]
\item Diffusion of ligands are assumed to be fast enough such that the reception is not mass transport limited. Therefore, the ligands are uniformly distributed in the reception space, and all of the receptors are always exposed to the same concentration of ligands.
\item Biological interference of any chemically similar molecule is neglected.
\end{inparaenum}
These assumptions are prevalent in MC and biosensor studies \cite{Akyildiz2013} \cite{Pierobon2011}, and lead to a pseudo-first order ligand-receptor dynamics, where the first time derivative of the mean number of bound receptors at time $t$ can be expressed by
\begin{equation}
\frac{d \overline{N_B (t)}}{dt} = k_+ c_L^R(t)(N_R - \overline{N_B(t)}) - k_- \overline{N_B(t)}, \label{eq:dNb/dt}
\end{equation}
where $k_+$ and $k_-$ are the intrinsic association and dissociation rate constants of the receptor-ligand complex, respectively. $N_R$ is the total number of receptors on the bioFET surface, and $c_L^R(t)$ is the ligand concentration in the vicinity of the recognition layer, i.e., in the reception space.

Since we assume reaction-limited operation for simplicity, we can neglect the transient phase between different levels of ligand concentration such that $c_L^R(t) = c_i$ for $t \in [t_i, t_i+1/B)]$, where $c_i$ is the ligand concentration corresponding to the $i$th message, $t_i$ is the transition time from the $(i-1)$th message to the $i$th message in reception space, and $1/B$ is the symbol duration with $B$ being the symbol transmission rate. In other words, biorecognition layer of the receiver is assumed to be exposed to a constant concentration $c_i$ for $t \in [t_i, t_i+1/B)]$.

Given the initial condition $\overline{N_B(t_i-\epsilon)} = N_{B,i-1}$ with $\epsilon \rightarrow 0$, the solution of \eqref{eq:dNb/dt} can be given as \cite{Berezhkovskii2013} \cite{Kuscu2015c}
\begin{multline}
\overline{N_B(t)} = \overline{N_{B,i}^{ss}} + \left( N_{B,i-1} - \overline{N_{B,i}^{ss}} \right) e^{-(k_+ c_i + k_-) (t-t_i)}  \\ \text{for}\ t \in [t_i, t_i+1/B),
\end{multline}
where $\overline{N_{B,i}^{ss}}$ is the mean number of bound receptors at steady-state, i.e., when $d\overline{N_B(t)}/dt = 0$. We can infer from this equation that although the ligand concentration level immediately changes in the reception space, it takes a certain time for the receptors to adapt the concentration level of the new message. The time to reach steady-state is governed by the reaction timescale $\tau_B = (k_+ c_i + k_-)^{-1}$; thus, for higher ligand concentrations, the adaptation time of the recognition layer decreases. The mean number of bound receptors at steady-state is given by
\begin{equation}
N_{B,i}^{ss} = \frac{k_+ c_i}{k_+ c_i + k_-} N_R = \frac{c_i}{c_i + K_D} N_R, \label{ss}
\end{equation}

Receiver is assumed to sample the state of receptors at steady-state; thus, the mean number of occupied receptors corresponding to the $i$th message can be given as $\overline{N_{B,i}} = \overline{N_{B,i}^{ss}}$. The occupation states of the receptors fluctuate around the mean value even in the steady-state. This phenomenon is referred to as receptor noise, which we discussed in Section \ref{Principles}. As being a summation of Bernoulli random variables, the number of bound receptors at steady-state follows the well-known binomial distribution, variance of which is given by
\begin{equation}
Var(N_{B,i}) = \frac{k_- k_+ c_i}{(k_+ c_i + k_-)^2} N_R. \label{eq:variance}
\end{equation}
The autocorrelation function for the stationary fluctuations of the number of bound receptors at steady-state can be approximated with a single exponential \cite{Bialek2005}:
\begin{equation}
R(\tau) = Var(N_{B,i}) e^{-\frac{\tau}{\tau_B}}, \label{eq:acf}
\end{equation}
where the characteristic timescale of ligand-receptor binding $\tau_B$ is also being the correlation time of binding noise. The Fourier Transform of \eqref{eq:acf} gives the Power Spectral Density (PSD) of the fluctuations:
\begin{equation}
S_{\Delta N_B}(f) = Var(N_{B,i}) \frac{2 \tau_B}{1+(2 \pi f \tau_B)^2}. \label{eq:SNb}
\end{equation}

\subsection{Model of Transducer Unit}
The charged ligands bound to the surface receptors induce opposite charges on the NW surface to some extent depending on their intrinsic charge. The mean amount of charge generated by the bound ligands for the $i$th message is given by $Q_i = \overline{N_{B,i}} \, N_e \, q_{eff}$, where $N_e$ is the average number of free electrons per ligand molecule. $q_{eff}$ is the mean effective charge induced by a single electron on a ligand molecule which is given in \eqref{qeff}, where the mean vertical distance of ligand electrons at the bound state to the transducer surface is assumed equal to the length of the receptor molecule $r = L_R$.
\begin{figure}[!t]
\centering
\includegraphics[width=7cm]{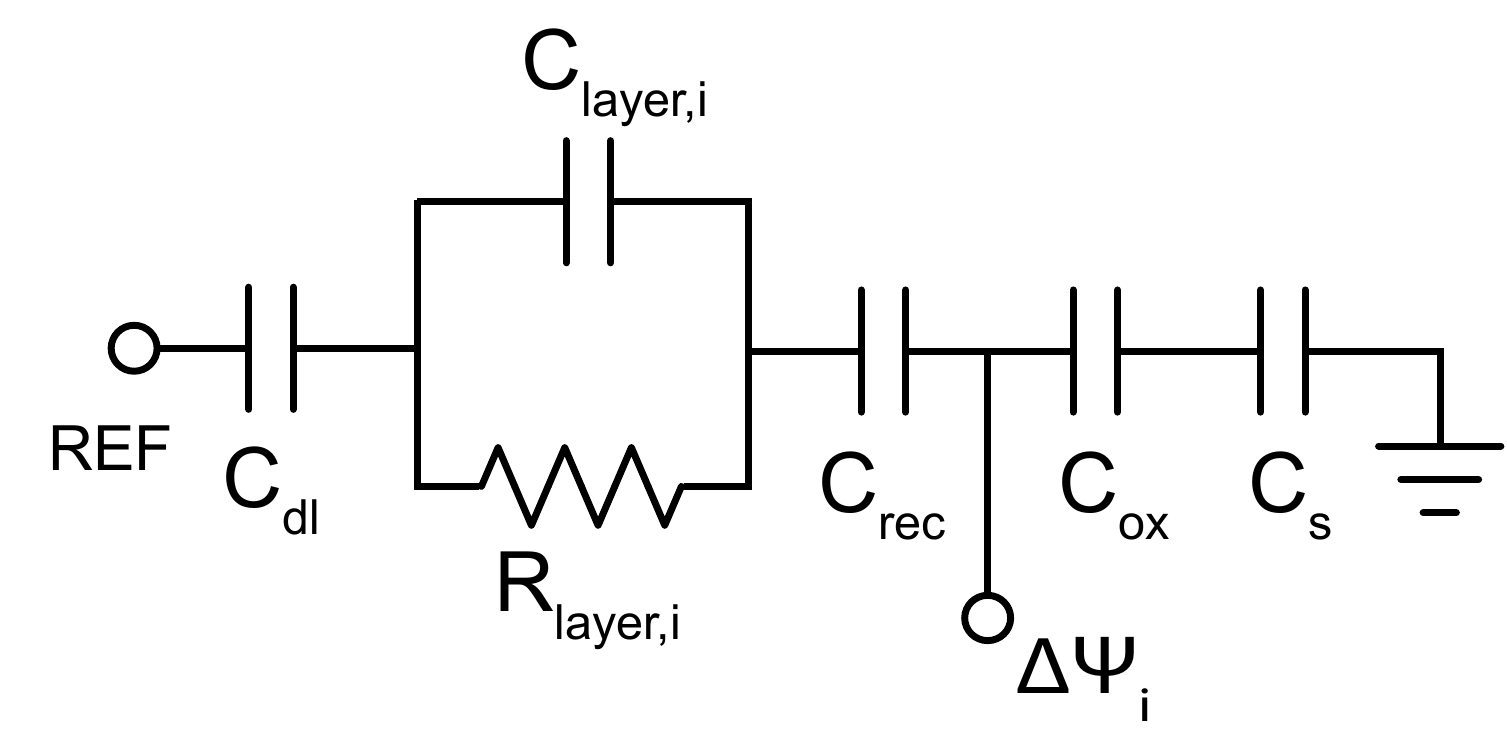}
\caption{Equivalent circuit model for the transducer of the SiNW FET-based MC receiver \cite{Deen2006} \cite{Kuscu2015c} \cite{Spathis2015}. \emph{REF} denotes the reference electrode, which stabilizes the gate voltage $V_G$.}
\label{fig:Circuit}
\end{figure}

The induced charges on the NW surface are translated into the variation of the surface potential through the equivalent circuit of the transducer \cite{Deen2006} \cite{Kuscu2015c} \cite{Spathis2015}, which is demonstrated in Fig. \ref{fig:Circuit}. By neglecting the current through $R_{layer}$, i.e., resistance of the layer of bound ligands, which is on the order of tens of $G \Omega$s \cite{Spathis2015} \cite{Bang2008}, the mean potential difference resulting from the bound ligands can be written as
\begin{equation}
\overline{\Delta \Psi_{i}} = \frac{Q_i}{C_{eq,i}},
\end{equation}
where $C_{eq,i}$ is the overall capacitance of the equivalent circuit:
\begin{multline}
C_{eq,i} = \left((C_{ox} W L)^{-1} + (C_{s} W L)^{-1}\right)^{-1}\\ + \left(C_{rec}^{-1} + C_{layer,i}^{-1} + (C_{dl} W L)^{-1}\right)^{-1}.
\end{multline}
Here, $C_{ox}$, $C_{s}$ and $C_{dl}$ are the silicon oxide (SiO$_2$), the semiconductor, i.e. SiNW, and the double layer capacitances per unit area, respectively; $C_{rec}$ and $C_{layer,i}$ are the capacitances of the receptor layer and the layer of bound ligands when the $i$th message is received, respectively; and $W$ and $L$ are the width and length of the transducer's active region. $C_{ox} = \epsilon_{ox}/t_{ox}$ with $\epsilon_{ox}$ and $t_{ox}$ being the permittivity and the thickness of the oxide layer. $C_{rec} = N_R \times C_{mol,R}$, $C_{layer,i} = \overline{N_{B,i}} \times C_{mol,L}$ with $C_{mol,R}$ and $C_{mol,L}$ being the capacitance of a single receptor and a single ligand molecule, respectively.
\begin{figure*}[!t]
 \centering
   \subfigure[]{
 \includegraphics[width=4.2cm]{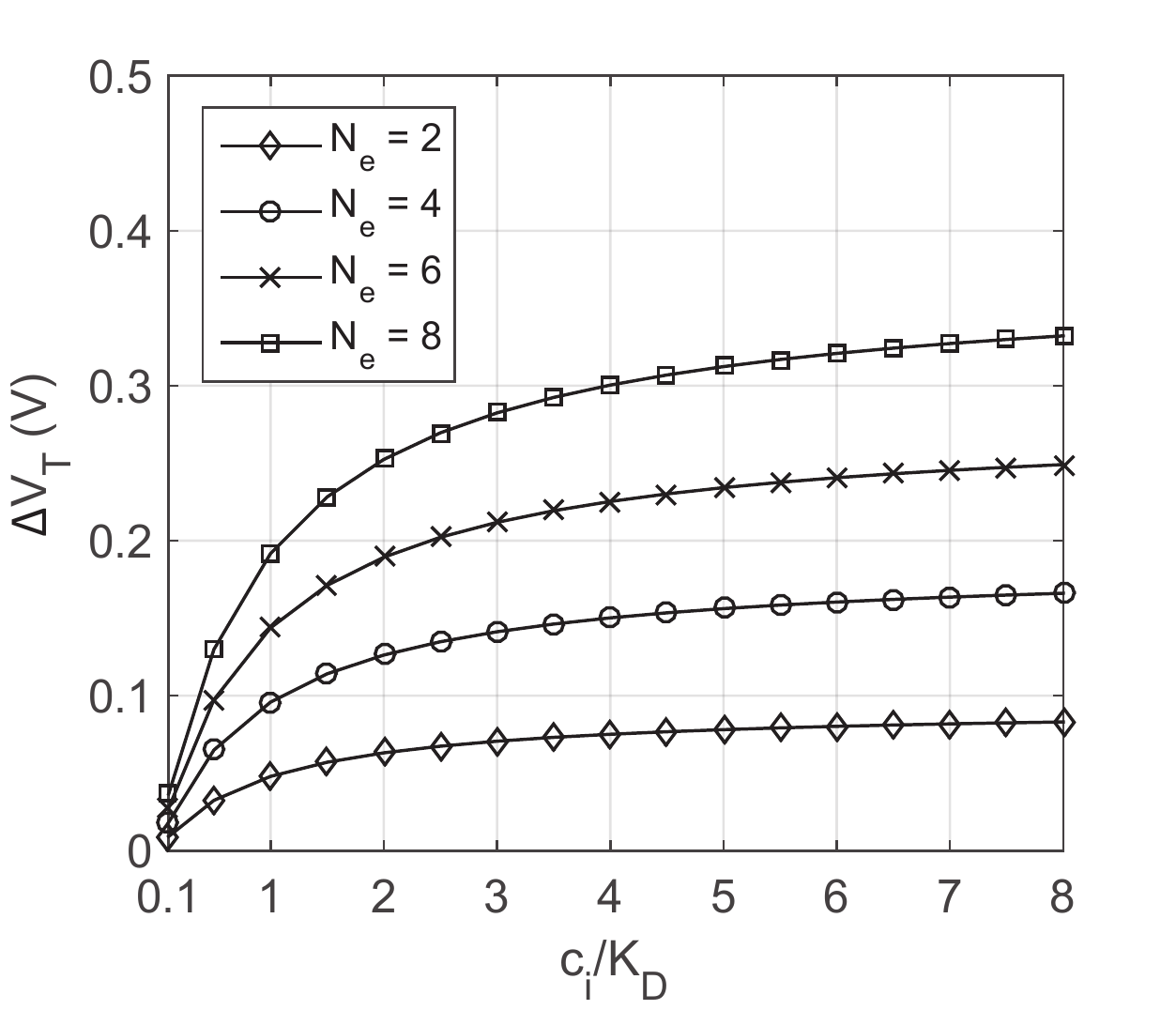}
 \label{fig:response_Ne}
 }
  \subfigure[]{
 \includegraphics[width=4.2cm]{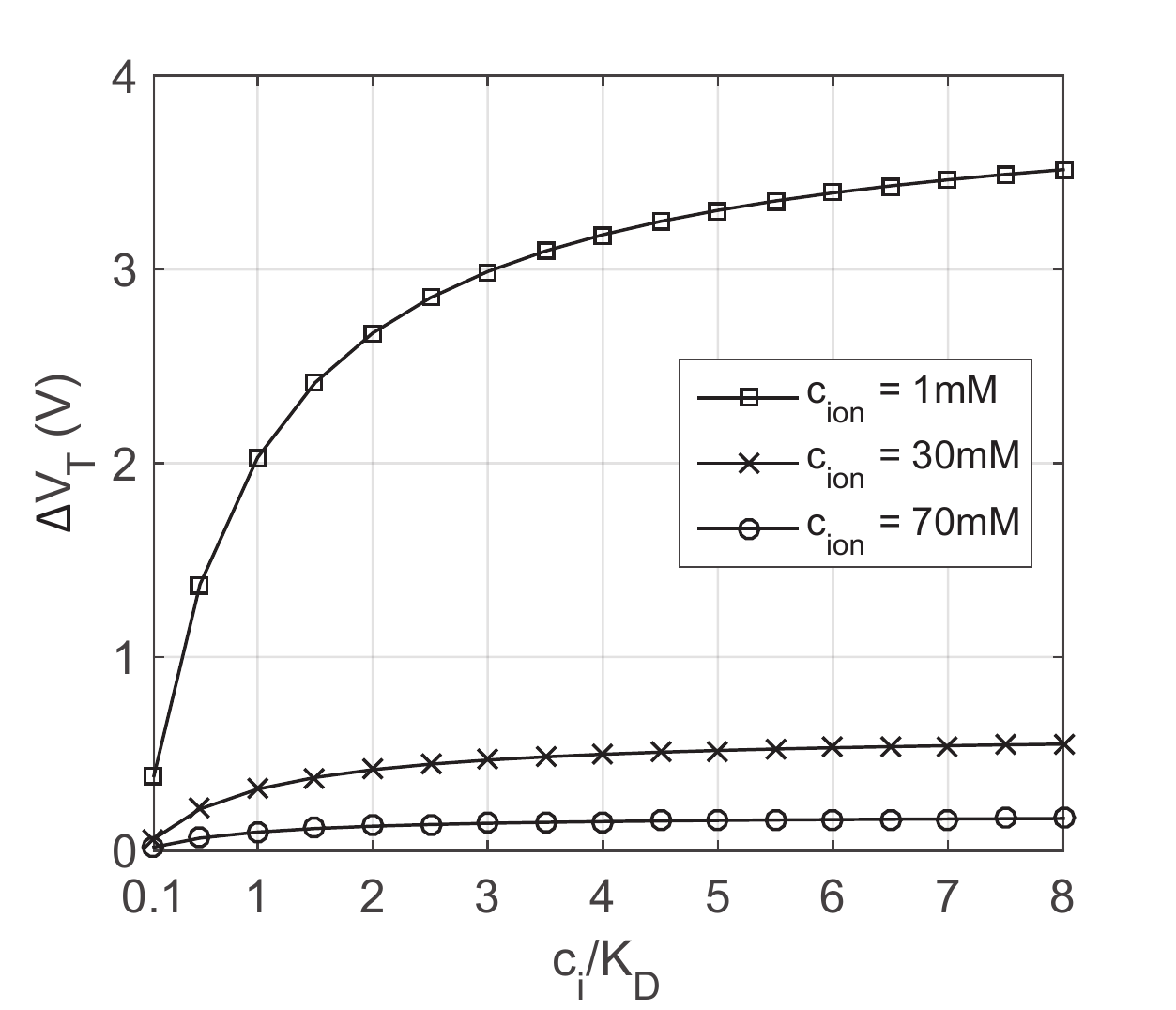}
 \label{fig:response_cion}
 }
    \subfigure[]{
 \includegraphics[width=4.2cm]{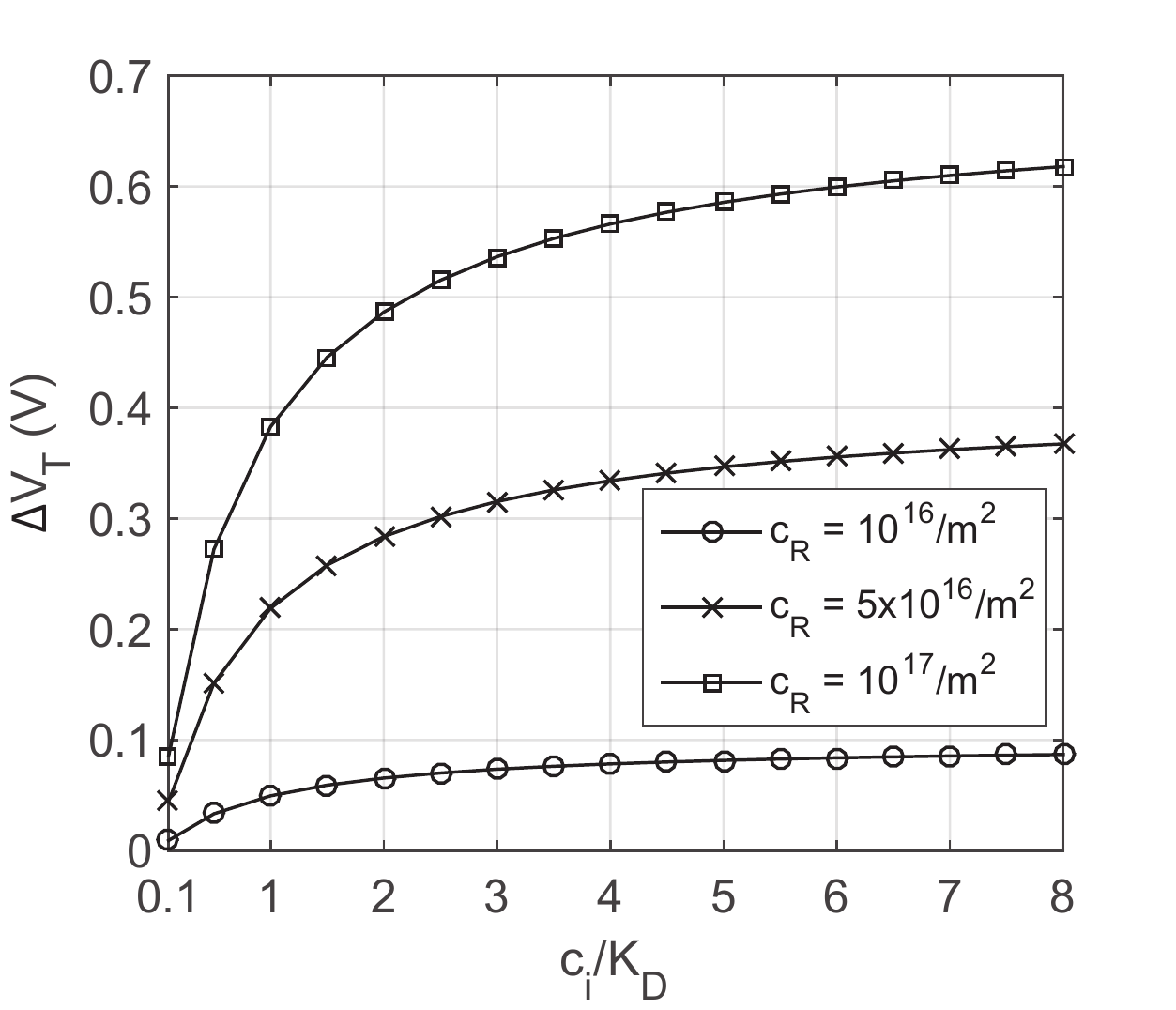}
 \label{fig:response_cR}
 }
  \subfigure[]{
 \includegraphics[width=4.2cm]{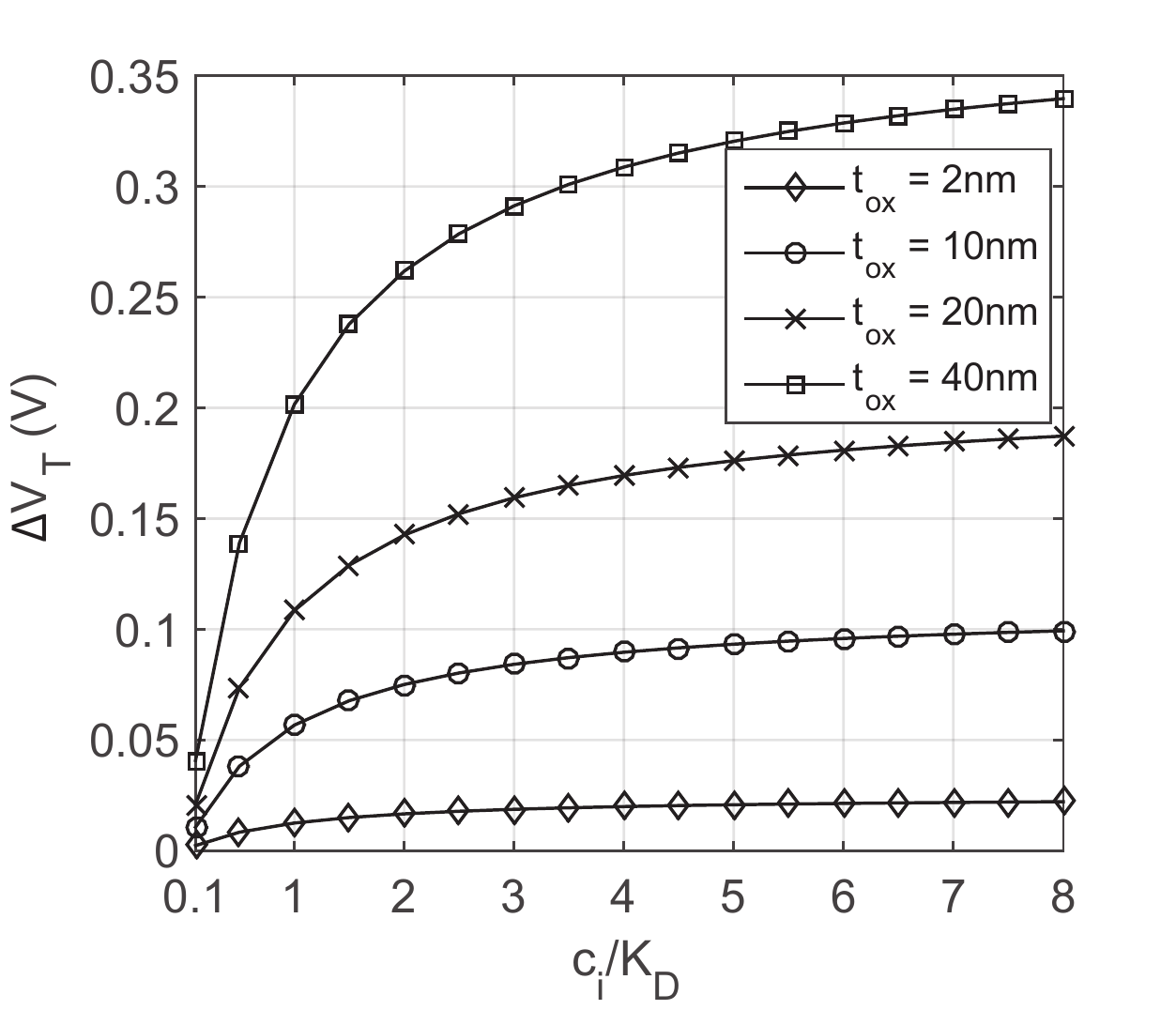}
 \label{fig:response_tox}
 }
 \caption{Receiver response in terms of difference in threshold voltage, i.e., $\Delta V_T = \Delta I_{DS}/g_m$, as a function of ligand concentration $c_i$ for (a) different mean number of electrons per ligand $N_e$, (b) different ionic concentration $c_{ion}$, (c) different concentration of surface receptors $c_R$ (note that $N_R = c_R W L$), (d) different oxide layer thickness $t_{ox}$.}
 \label{fig:response}
 \end{figure*}

\begin{table}[!b]\scriptsize
\centering
\caption{Simulation Parameters}
\begin{tabular}{ l | l }
   \hline \hline
   Size of active region ($W \times L$) & $0.1 \times 5$ ($\mu m$) \\ \hline
   Temperature ($T$) & $298$ ($K$) \\ \hline
   Relative permittivity of SiO$_2$ layer ($\epsilon_{ox}/\epsilon_0$) & $3.9$ \\ \hline
   Thickness of SiO$_2$ layer ($t_{ox}$) & $17.5$ ($nm$) \\ \hline
   Effective mobility ($\mu_{eff}$) & $16 \times 10^{-3}$ ($m^2 V^{-1} s^{-1}$) \\ \hline
   Drain-source voltage ($V_{DS}$) & $0.1$ ($V$) \\ \hline
   Relative permittivity of solvent ($\epsilon_R/\epsilon_0$) & $78$ \\ \hline
   Ionic concentration of medium ($c_{ion}$) & $70$ ($mM$) \\ \hline 
   Trap density ($N_t$) & $2.3 \times 10^{24}$ ($eV^{-1} m^{-3}$) \\ \hline
   Tunneling distance ($\lambda$) & $0.05$ ($nm$) \\ \hline
   Average net charge of ligands ($N_e$) & $4$ \\ \hline  
   Length of receptor ($L_R$) & $4$ ($nm$) \\ \hline   
   Binding rate ($k_+$) & $2 \times 10^{-18}$ ($m^3 s^{-1}$) \\ \hline
   Unbinding rate ($k_-$) & $10$ ($s^{-1}$) \\ \hline
   Ligand concentration in reception space ($c_i$) & 4$K_D$  \\ \hline
   Concentration of receptors on the surface ($c_R$) & $2 \times 10^{16}$ ($m^{-2}$) \\ \hline
   Molecular capacitance ($C_{mol,L}, C_{mol,R}$) & $2 \times 10^{-20}$ ($F$) \\ \hline
   Capacitance of dielectric layer ($C_{dl}$) & $5 \times 10^{-2}$ ($F/m^2$)\\ \hline
   Capacitance of silicon ($C_s$) & $2 \times 10^{-3}$ ($F/m^2$) \\ \hline
\end{tabular}
\label{table:parameters}
\end{table}

The deviation of the surface potential is reflected into a change in the FET device threshold as $\overline{\Delta V_{TH,i}} = \overline{\Delta \Psi_i}$. We know that the drain-to-source current of FETs in the linear operation regime can be written as
\begin{equation}
I_{DS} = \frac{W}{L} \mu_{eff} C_{ox} (V_{GS} - V_{TH}) V_{DS}, \label{eq:Is}
\end{equation}
where $\mu_{eff}$ is the effective carrier mobility in the transducer channel, and $V_{DS}$ is the drain-to-source voltage, which is held constant \cite{Rajan2013-2}, $V_{GS}$ is the gate-to-source voltage, which is stabilized by the reference electrode. In case the threshold voltage is shifted, the mean variation reflected to $I_{DS}$ can be given by
\begin{equation}
\overline{\Delta I_{DS,i}} = \overline{\Delta V_{TH,i}} \frac{W}{L} \mu_{eff} C_{ox} V_{DS}, \label{eq:Is}
\end{equation}
We can simply express the mean current variation by $\overline{\Delta I_{DS,i}} = g_m \overline{\Delta V_{TH,i}} = g_m \Delta \Psi_i$, where $g_m$ is the transconductance of the device, which can be given as
\begin{equation}
g_m = \frac{W}{L} \mu_{eff} C_{ox} V_{DS}, \label{eq:conductance}
\end{equation}
The processor unit of the receiver uses the signal $\Delta I_{DS,i}$, to infer the incoming ligand concentration $c_i$; and thus the molecular message $i$, based on a predefined CSK scheme.

As discussed in Section \ref{Principles}, random diffusion of free electrons on the layer of bound ligands results in thermal noise. The uncertainty in the location of electrons are reflected to fluctuations in the threshold voltage of bioFET antenna. Using the thermal noise model derived in \cite{Spathis2015}, the PSD of voltage fluctuations on the layer is given by $S_{\Delta V_{TH,i}^{T,layer}} = 4 k_B T R_{layer,i}$, where $R_{layer,i}$ is the mean resistance of the layer when the $i$th message is received. The fluctuations are reflected to the threshold voltage through the RC filter given in Fig. \ref{fig:Circuit}. Using the its transfer function, the PSD of the thermal noise contribution on $\Delta V_{TH,i}$ is written as \cite{Kuscu2015c} \cite{Spathis2015}

\small
\begin{equation}
S_{\Delta V_{TH,i}^T}(f) = S_{\Delta V_{TH,i}}^{T,layer} \left( 1 + \left[2 \pi R_{layer,i} C_{eq}' f\right]^2 \right)^{-1}, \label{eq:thermalpsd}
\end{equation}\normalsize
with \small $C_{eq}' = C_{layer,i}+\left[(C_{dl}^{-1}+C_{ox}^{-1}+C_s^{-1}) (WL)^{-1}+C_{rec}^{-1}\right]^{-1}.$ \normalsize
In addition to the receptor and thermal noise, the low-frequency operation of bioFET-based molecular antenna is also suffered from $1/f$ noise. Using the well-known number fluctuation model \cite{Kuscu2015c} \cite{Rajan2013-2}, the PSD of $1/f$ noise can be written as follows
\begin{equation}
S_{\Delta V_{TH}^F}(f) = \frac{\lambda k T q^2 N_t}{W L C_{ox}^2 |f|},
\end{equation}
where $\lambda$ is the characteristic tunneling distance, and $N_t$ is the trap density of the NW channel. $1/f$ noise is independent of the received signals, and shows an additive behavior on the overall threshold voltage fluctuations \cite{Spathis2015}.

\subsection{Performance Analysis}
In this section, we analyze the receiver performance in terms of expected response, sensitivity and SNR. Table \ref{table:parameters} lists the default parameter values used in the analysis. Considering that most of the nanonetwork applications are envisioned for \textit{in vivo} scenarios, the parameter values pertaining to medium characteristics are selected based on the physiological conditions. For example, the default value of ionic concentration, $c_{ion}$, is selected according to the reported value for bovine serum (70mM) \cite{Okada1990}; but in the sensitivity and SNR analyses, we investigate the performance for ionic concentrations corresponding to a wide range of solutions including highly diluted solutions ($c_{ion}$=1mM) and human blood plasma ($c_{ion}>$100mM) \cite{Duan2012}. The solvent is water by default in physiological conditions, which implies the relative permittivity, $\epsilon_R/\epsilon_0 = 78$ \cite{Hediger2012}. The employed receptors on the FET surface are considered to be aptamers, the production process of which provides full control over the selection of length, binding and unbinding rates, as well as the type of corresponding ligand molecules (see Section \ref{Receptors}). The default length of receptors is set to 4nm, which corresponds to 12 base pair-aptamers. Binding and unbinding rates, $k_+$ and $k_-$, are set, considering the accepted values in the MC literature \cite{Pierobon2011} and the range of rates that aptamers can provide \cite{Luzi2003}. Aptamers can bind to a large set of ligands, such as, aptamers, small proteins, RNA and DNA, and even non-organic molecules, which can attain a broad range of elementary charges, as will be discussed in Section \ref{ReceiverResponse}. The capacitance of receptors and ligands, $C_{mol,R}$ and $C_{mol,L}$, are selected based on the reported values for oligonucleotides and small molecules \cite{Lu2008}, which correspond to a mean value of $2 \times 10^{-20}$F. The relative permittivity of SiO$_2$ layer is reported as $\epsilon_{ox}/\epsilon_0 = 3.9$ \cite{Deen2006}. The thickness of the SiO$_2$ layer, $t_{ox}$, is a design parameter, for which we select a default value of $17.5$nm corresponding to the design in \cite{Deen2006}. However, we carry out the analyses for a wide range of $t_{ox}$ values considering that the thickness can be reduced to below $5$nm, as reported in \cite{Kobayashi1998}. Depending on the fabrication, the tunneling distance for SiO$_2$ is on the order of 0.01-0.1nm \cite{Rajan2011a}. We set $\lambda = 0.05$nm as reported in \cite{Deen2006}, which also reports the effective mobility as $16 \times 10^{-3}$m$^2/$Vs. The capacitance of dielectric layer, $C_{dl}$, and silicon substrate, $C_s$ are design parameters for which we set the default values as reported in \cite{Deen2006}. The trap density $N_t$ depends on the quality of fabrication, and can attain a wide range of values \cite{Rajan2011a}. As default, we set a moderate value of $2.3 \times 10^{24}$eV$^{-1}$m$^{-3}$. In the analyses, we also discuss the experimental results obtained in \cite{Duan2012} and compare them with our numerical results.

\subsubsection{Receiver Response}
\label{ReceiverResponse}
Fig. \ref{fig:response_Ne}, presents the receiver response in terms of the expected variation of $\Delta V_T$, which is the mean variation of the channel current normalized by the device transconductance $g_m$, as a function of the average number of elementary charges that ligands possess. As is seen, the net charge of ligands critically affects the receiver response, since the transducer operation is based on the field effect generated by the ligand charges. As reviewed in Section \ref{Receptors}, oligonucleotides are negatively charged in physiological conditions, i.e., at pH 7.4, due to their highly charged phosphate backbone. For example, a DNA sequence with 4 base-pairs at pH 7.4 can attain a net charge of $-8e$. Likewise, small proteins and antigens, depending on the pH of the environment, can attain a net charge of up to $\pm 4$e \cite{Tsai2006}.

Employing highly charged ligands as information molecules is not sufficient to obtain a proper receiver performance, since the ligand charges are substantially screened by the ions present in the surrounding medium. The major determinant of this so-called Debye screening is the ionic concentration of the medium, i.e., $c_{ion}$. In Fig. \ref{fig:response_cion}, we present the receiver response for different values of $c_{ion}$. As it is clear, the shift in the threshold voltage significantly increases with decreasing ionic concentration. 20-fold increase in the threshold voltage is obtained when the ions in the solution is diluted to 1mM from 70mM concentration, which is the ionic concentration of bovine serum \cite{Okada1990}. Note that physiological conditions for human body generally imply ionic concentrations higher than 100mM \cite{Rajan2013}.
\begin{figure}[!h]
 \centering
   \subfigure[]{
 \includegraphics[width=6cm]{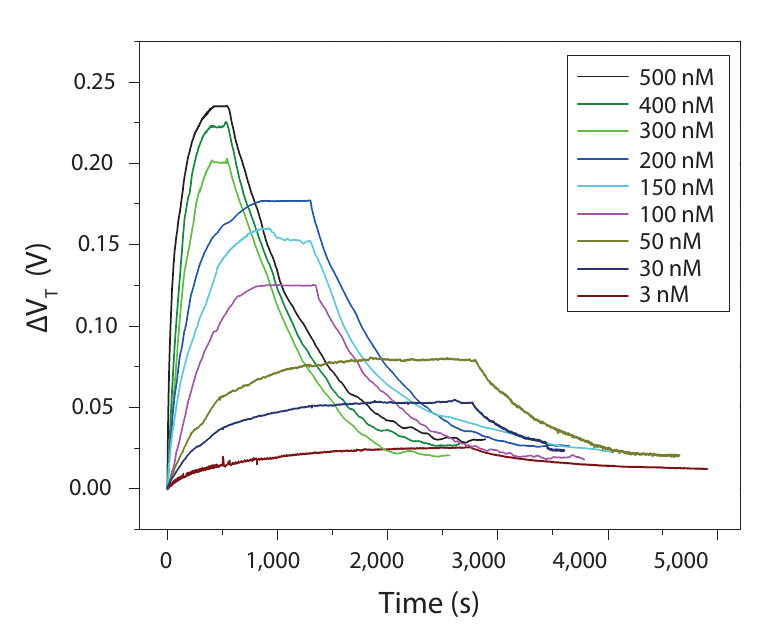}
 \label{fig:exp1}
 }
  \subfigure[]{
 \includegraphics[width=6cm]{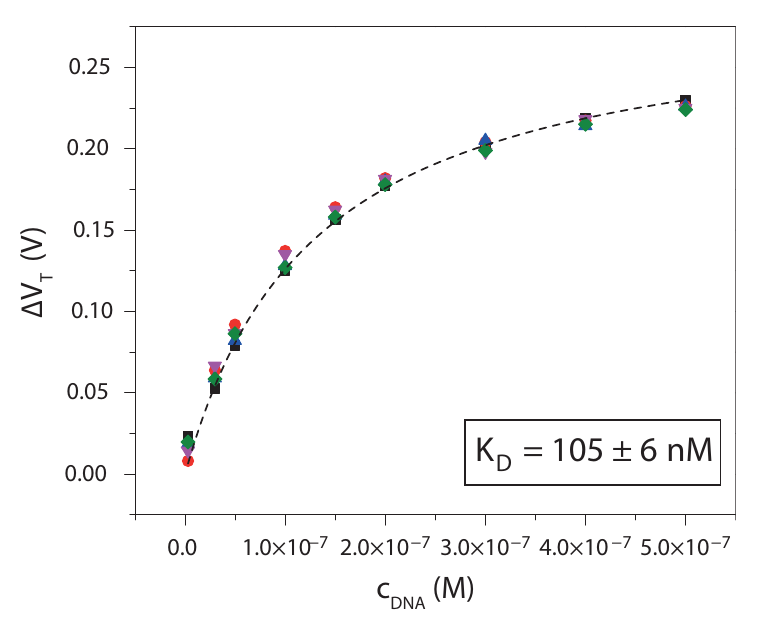}
 \label{fig:exp2}
 }
 \caption{The results of experiments conducted in \cite{Duan2012} for the detection of DNA concentration by a SiNW FET functionalized with HMGB1 proteins: (a) Threshold voltage $\Delta V_T$ vs. time data for the detection of DNA with HMGB1 receptors on the same SiNW bioFET device for different concentrations of DNA. Each DNA delivery is followed by 1 M NaCl buffer to accelerate the desorption of bound DNAs from the surface. (b) $\Delta V_T$ as a function of DNA concentration. The dissociation constant $K_D$ for DNA-HMGB1 ligand-receptor pair is determined by the authors as $105 \pm 6$ nM. Adapted by permission from Macmillan Publishers Ltd: Nature Nanotechnology, \cite{Duan2012}, copyright 2012.}
 \label{fig:exp}
 \end{figure}
Third analysis is performed for varying concentration of surface receptors. As is seen from Fig. \ref{fig:response_cR}, deploying receptor molecules more densely on the surface of the semiconductor channel, increases the receiver response almost linearly. However, size restrictions and possible correlations among densely deployed receptors, which are not captured by this model, should be accounted for in a real world implementation.
Lastly, we analyze the effect of oxide layer thickness $t_{ox}$, which is the main determinant of the oxide capacitance $C_{ox}$, on the receiver response. Fig. \ref{fig:response_tox} demonstrates the results for conventional values of $t_{ox}$. As can be inferred, higher $t_{ox}$ implies a better receiver performance.
\begin{figure*}[!t]
 \centering
   \subfigure[]{
 \includegraphics[width=4.2cm]{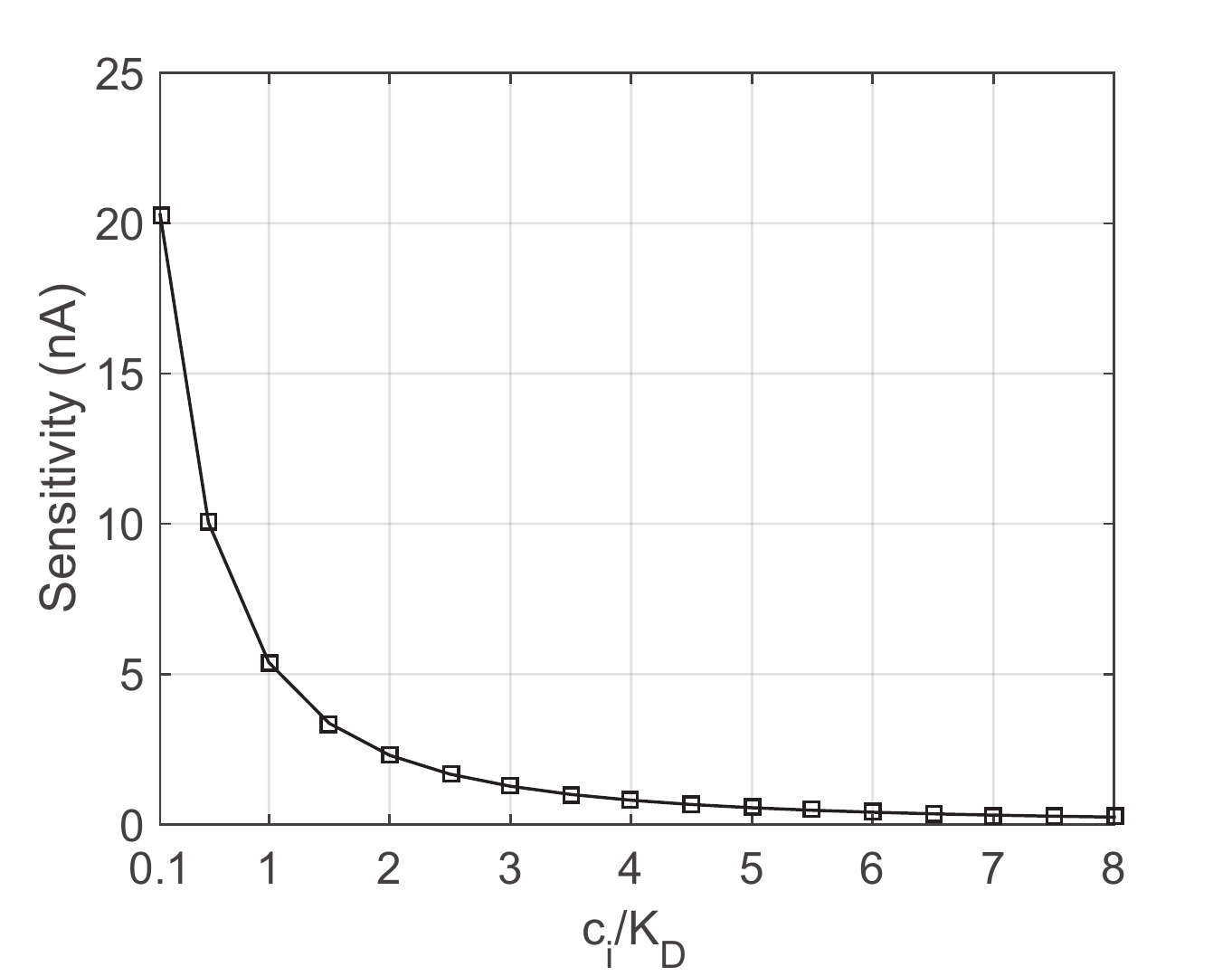}
 \label{fig:SENc}
 }
  \subfigure[]{
 \includegraphics[width=4.2cm]{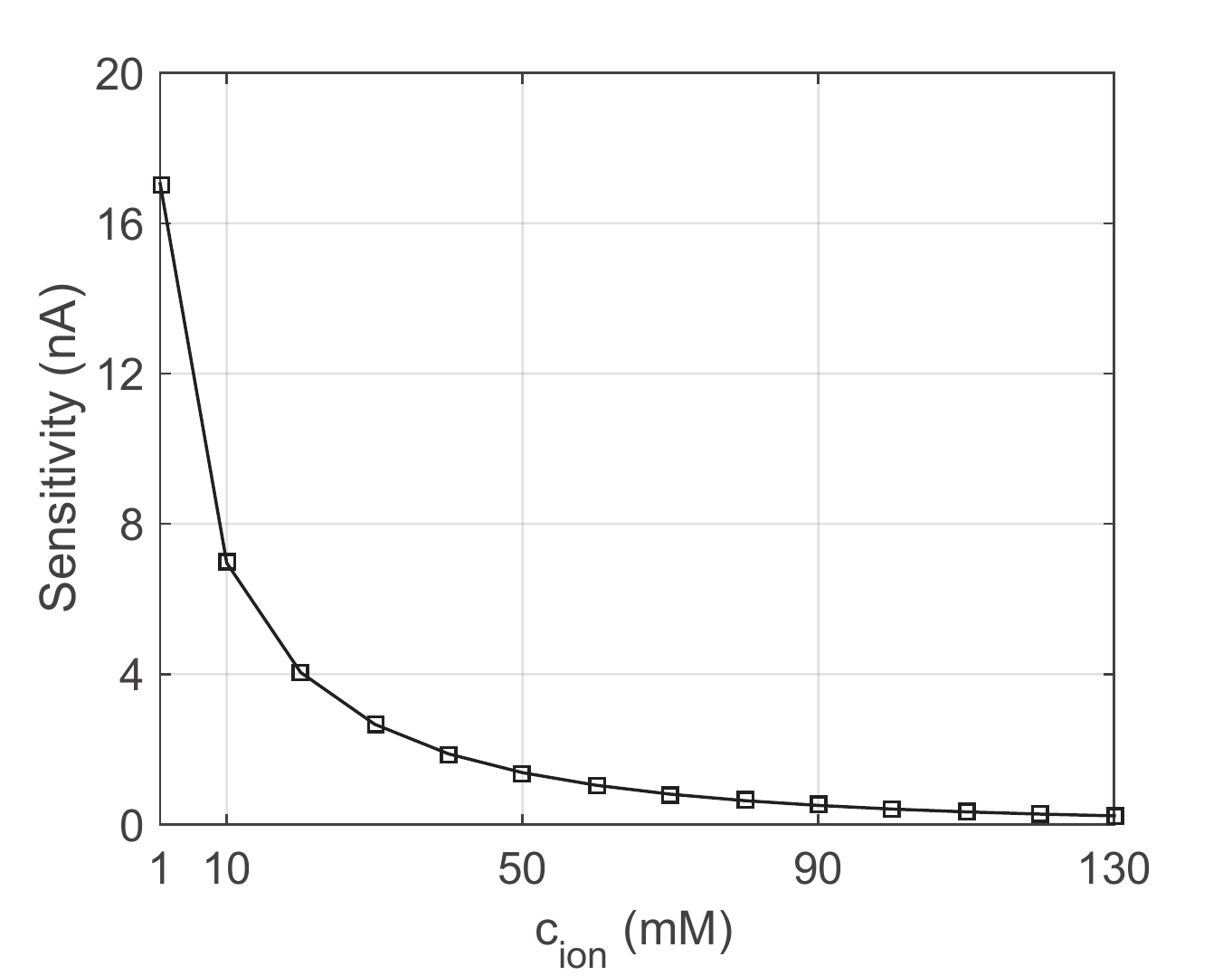}
 \label{fig:SENcion}
 }
    \subfigure[]{
 \includegraphics[width=4.2cm]{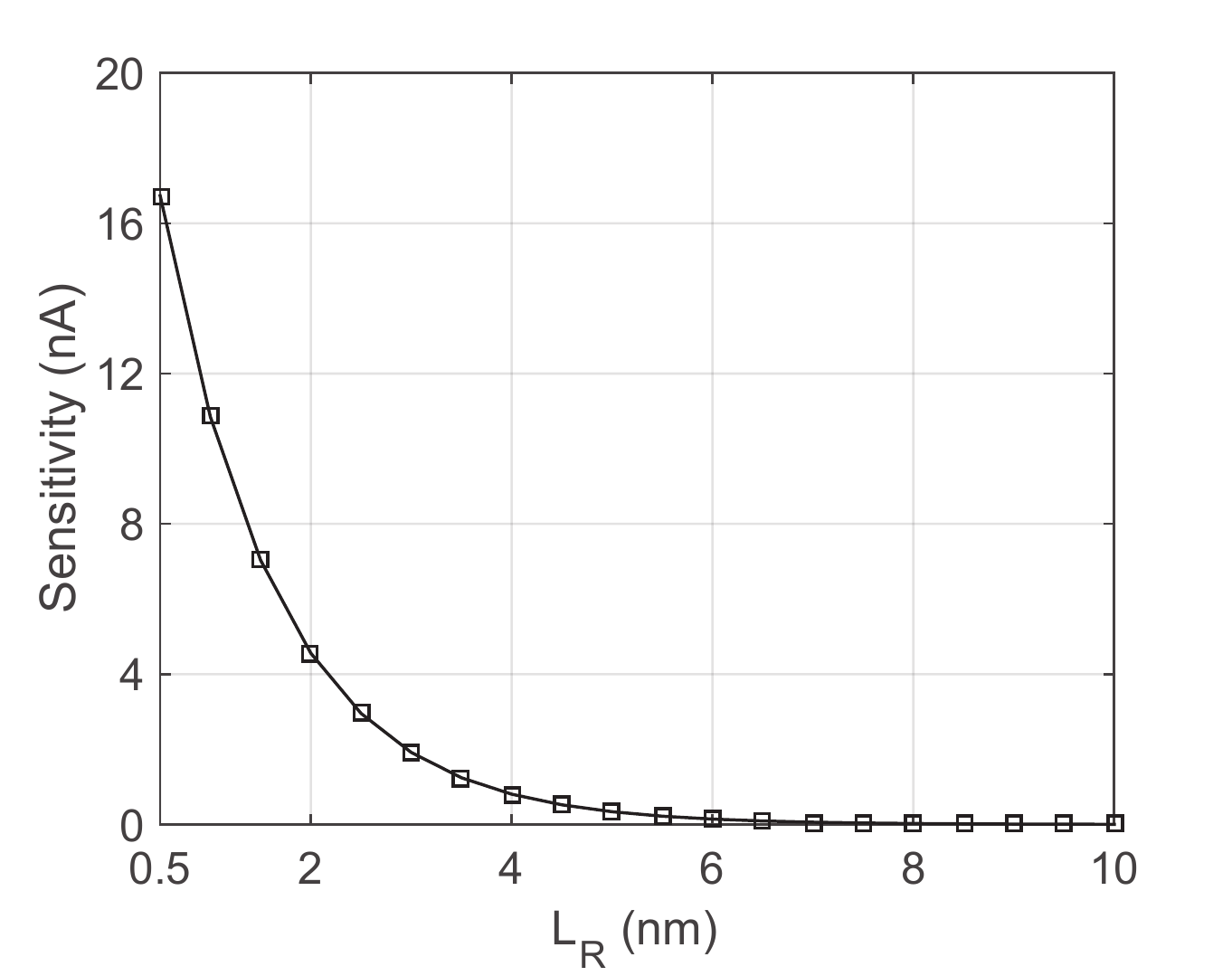}
 \label{fig:SENLr}
 }
  \subfigure[]{
 \includegraphics[width=4.2cm]{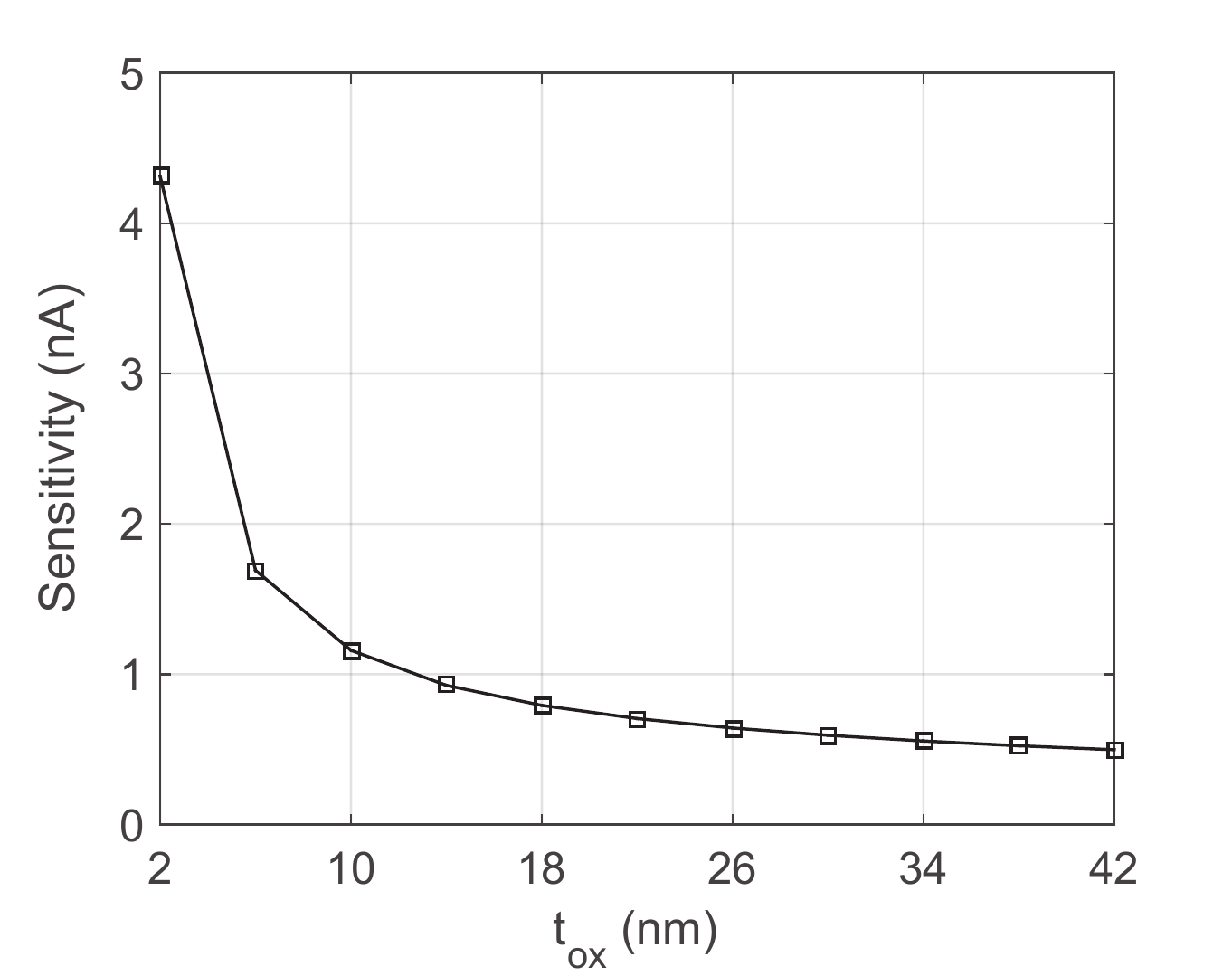}
 \label{fig:SENtox}
 }
 \caption{Receiver sensitivity with input concentration $c_i$ normalized by $K_D$. Sensitivity with (a) varying concentration $c_i$ of information-carrying ligands, (b) varying ionic concentration $c_{ion}$, (c) varying receptor length $L_R$, and (d) varying oxide layer thickness $t_{ox}$.}
 \label{fig:SEN}
 \end{figure*}

The response of a typical SiNW bioFET, which is the base of the proposed receiver, to varying ligand concentration can also be observed from the experimental results presented in Fig. \ref{fig:exp}. These results have been acquired from \cite{Duan2012}, where the authors employed a SiNW bioFET to quantify the affinity of DNA sequences with HMGB1 proteins. HMBG1 proteins of an unspecified concentration were deployed on the SiNW surface, and varying concentration of DNAs as the charged ligands at pH 7.4 were introduced into the reception space of the bioFET through a microfluidic system. The association on the surface through the ligand-receptor binding dynamics as the DNAs were supplied at a constant rate with different concentrations, and the dissociation after the steady-state was reached, can be observed from Fig. \ref{fig:exp1}. The binding dynamics of DNA and HMGB1 were confirmed to be reaction-limited; thus, it does not violate the well-mixed assumption made for the analytical model of biorecognition layer. Unfortunately, many of the critical system parameters, such as the concentration of the surface receptors, the average ligand charge, the average receptor length, each of which substantially affects the device response, were not provided in that study. Thus, we are not able to make a one-to-one comparison of the response given in Fig. \ref{fig:exp2} with the numerical results obtained using the model provided in this paper. Nevertheless, the experimental results clearly show the same saturation trend observed for the device response and very well fall into the range of the numerical results obtained in our analysis.

\subsubsection{Sensitivity Analysis}
Sensitivity measures the responsivity of the receiver to the varying concentration of ligands, thus, determines how successful the receiver is in discriminating the messages encoded into different ligand concentrations. It can be defined as the derivative of \eqref{eq:Is} with respect to input concentration $c_i$:

\small
\begin{equation}
S(c_i) = \frac{K_D N_e q_{eff} g_m  (N_R C_{mol,L} C_{eq,i} C_{p,i}^2 -1)}{(K_D + c_i)^2 C_{mol,L} C_{eq,i}^2 C_{p,i}^2}, \label{eq:sensitivity}
\end{equation}
\normalsize
where \small $C_{p,i} = C_{rec}^{-1} + C_{layer,i}^{-1} + (C_{dl} W L)^{-1}$. \normalsize Note that $C_{eq,i}$ and $C_{p,i}$ are functions of input concentration $c_i$.

Fig. \ref{fig:SEN} demonstrates the sensitivity with the input concentration normalized by the dissociation constant $K_D$, such that the normalized sensitivity in the results correspond to the amount of increase in the output current due to a unit increase in the normalized concentration $c_i/K_D$. As can be inferred from Fig. \ref{fig:SENc}, the sensitivity substantially decreases as the concentration level in the reception space increases. This is because the receptors on the biorecognition layer are saturated at higher ligand concentrations, thus become insensitive to the changes in the input. The results presented in Fig. \ref{fig:SENcion}  and Fig. \ref{fig:SENLr} reveal the effect of Debye screening such that the sensitivity decreases with increasing ion concentration and with increasing receptor length. Since it is not always possible to control the ionic concentration of the medium, it is critical to employ surface receptors with lengths smaller than the Debye length to obtain a highly sensitive receiver antenna. Physiological conditions generally imply ionic concentrations higher than $100mol/m^3$, which decrease the Debye length below $1nm$. As discussed in Section \ref{Receptors}, aptamers and natural receptors with smaller sizes are advantageous compared to antibodies to be employed as receptors in \emph{in vivo} applications. The last analysis in Fig. \ref{fig:SENtox} demonstrates that the thickness of the oxide layer $t_{ox}$ has also a significant effect on the sensitivity. As $t_{ox}$ increases, the oxide capacitance $C_{ox}$ decreases, which directly reduces the transconductance $g_m$ as given in \eqref{eq:conductance}. Fortunately, control over the oxide layer thickness is possible, and $t_{ox}$ values lower than $5nm$ are reported in the literature \cite{Kobayashi1998}.
\begin{figure*}[!t]
 \centering
   \subfigure[]{
 \includegraphics[width=4.2cm]{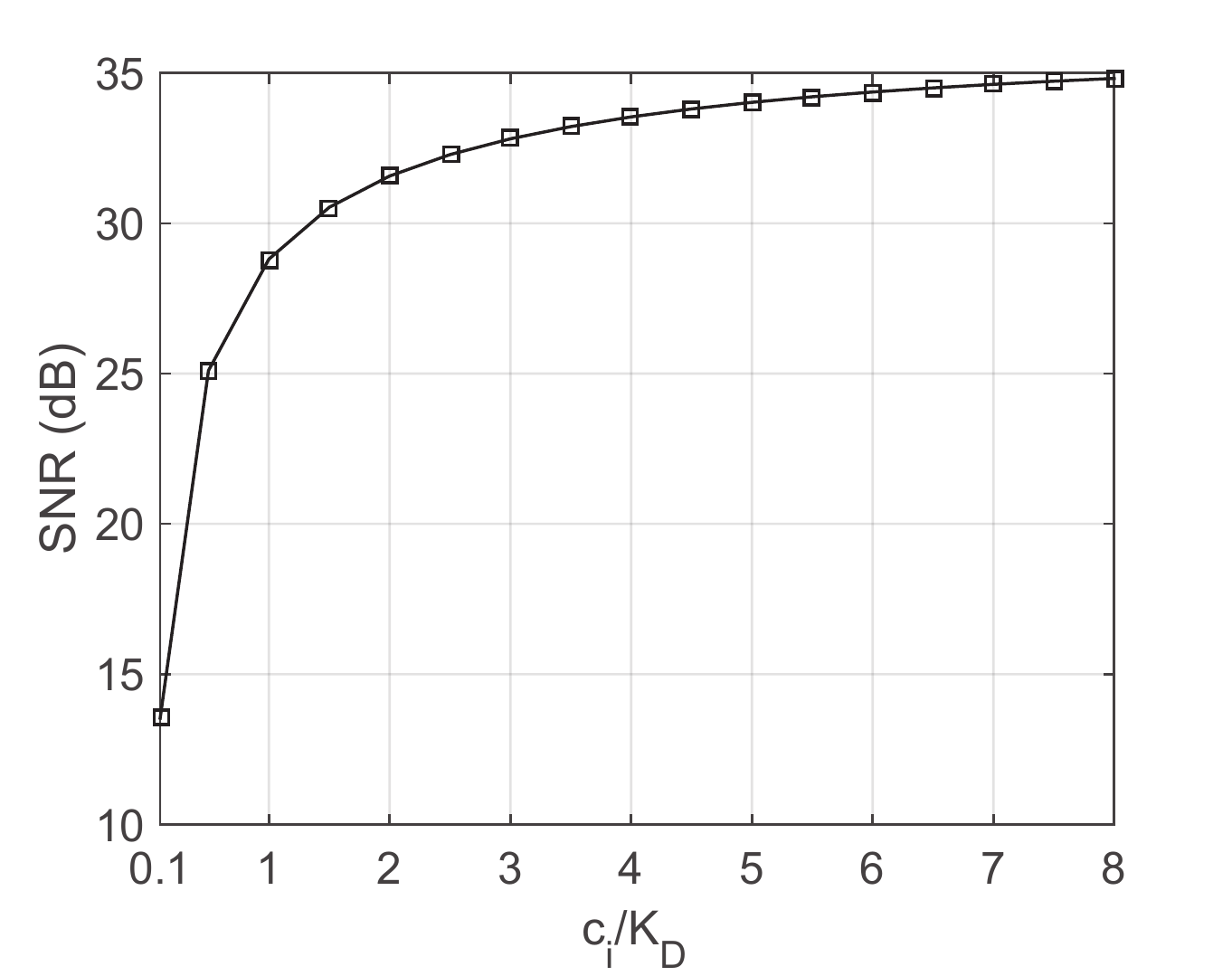}
 \label{fig:SNRc}
 }
  \subfigure[]{
 \includegraphics[width=4.2cm]{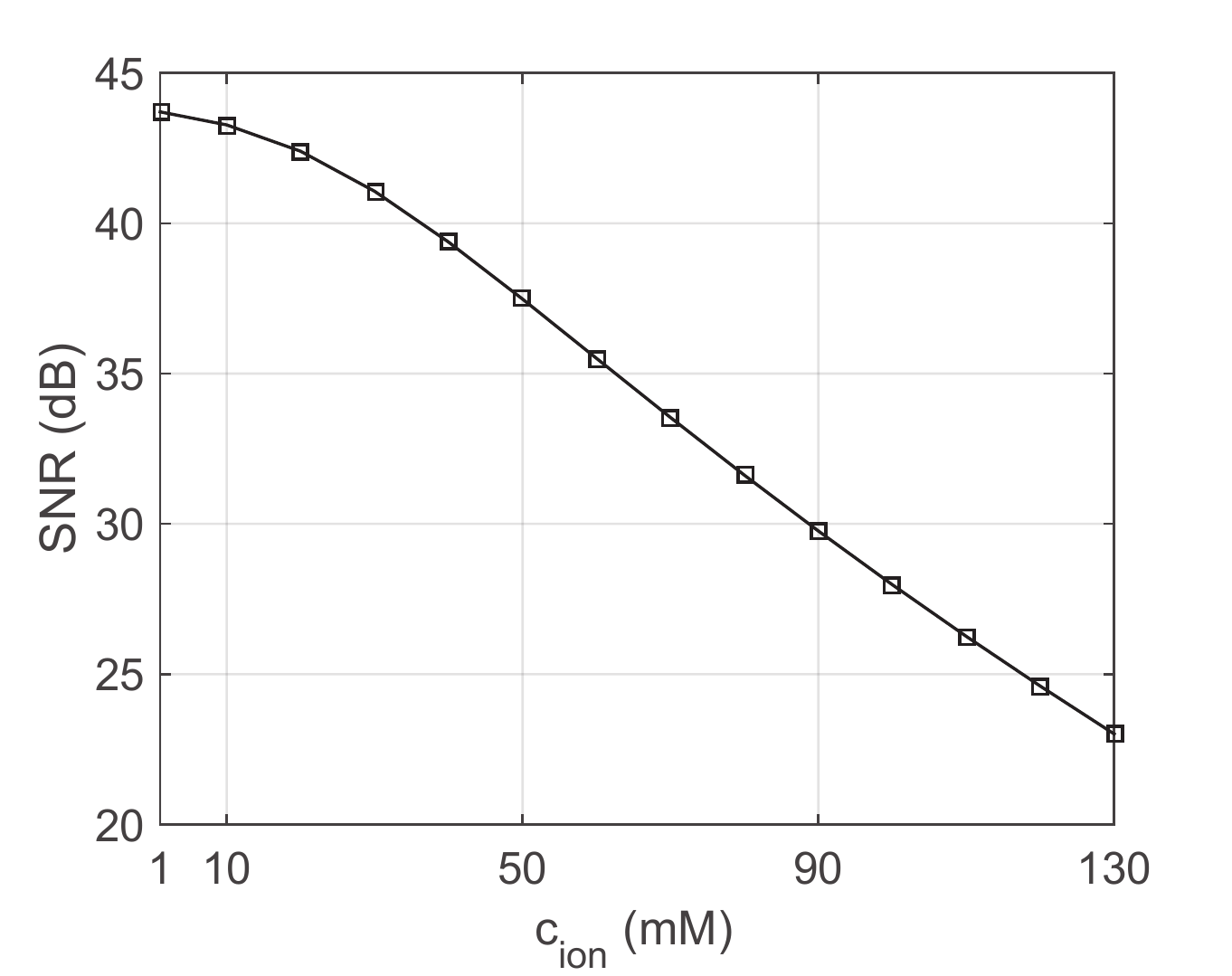}
 \label{fig:SNRcion}
 }
    \subfigure[]{
 \includegraphics[width=4.2cm]{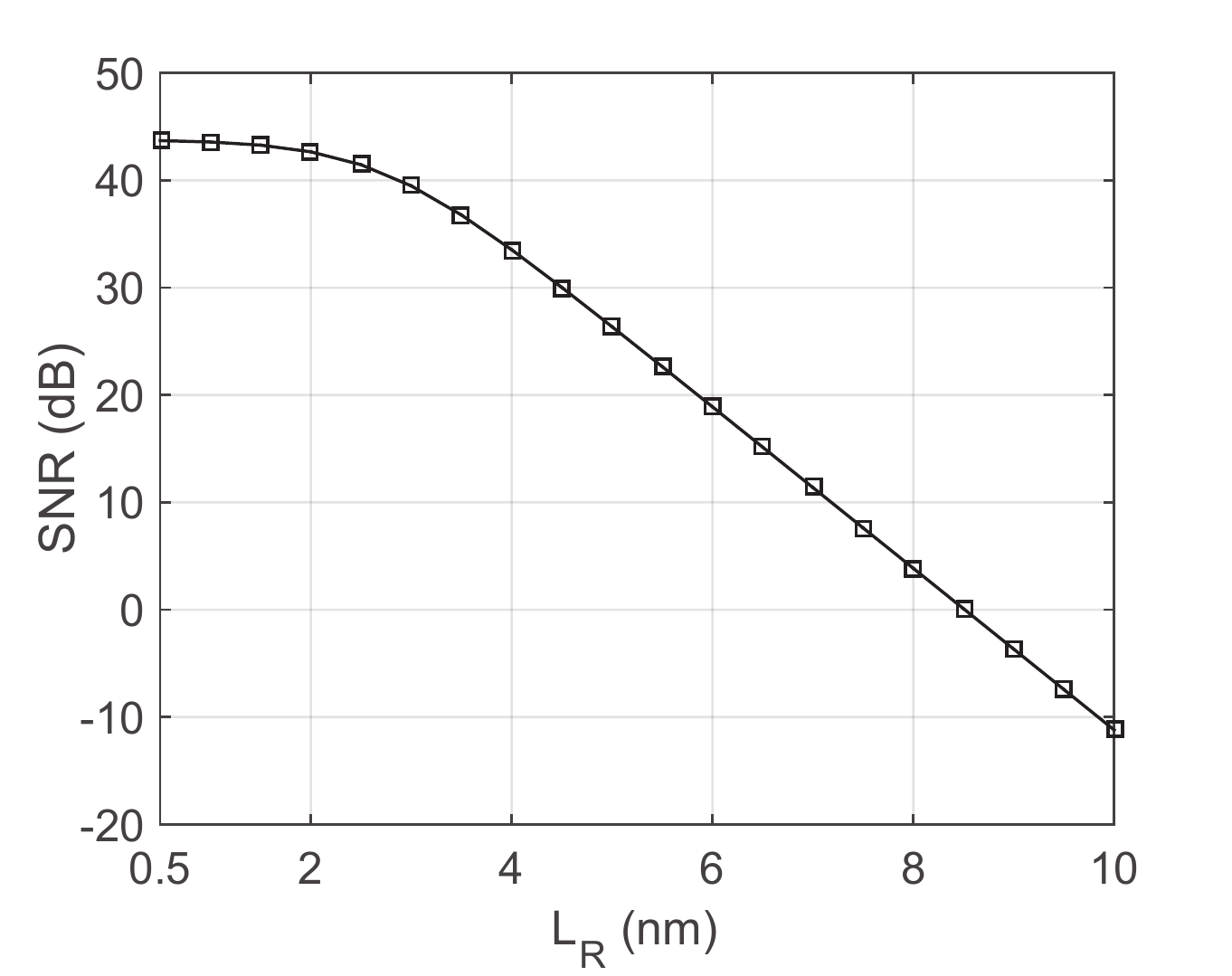}
 \label{fig:SNRLr}
 }
  \subfigure[]{
 \includegraphics[width=4.2cm]{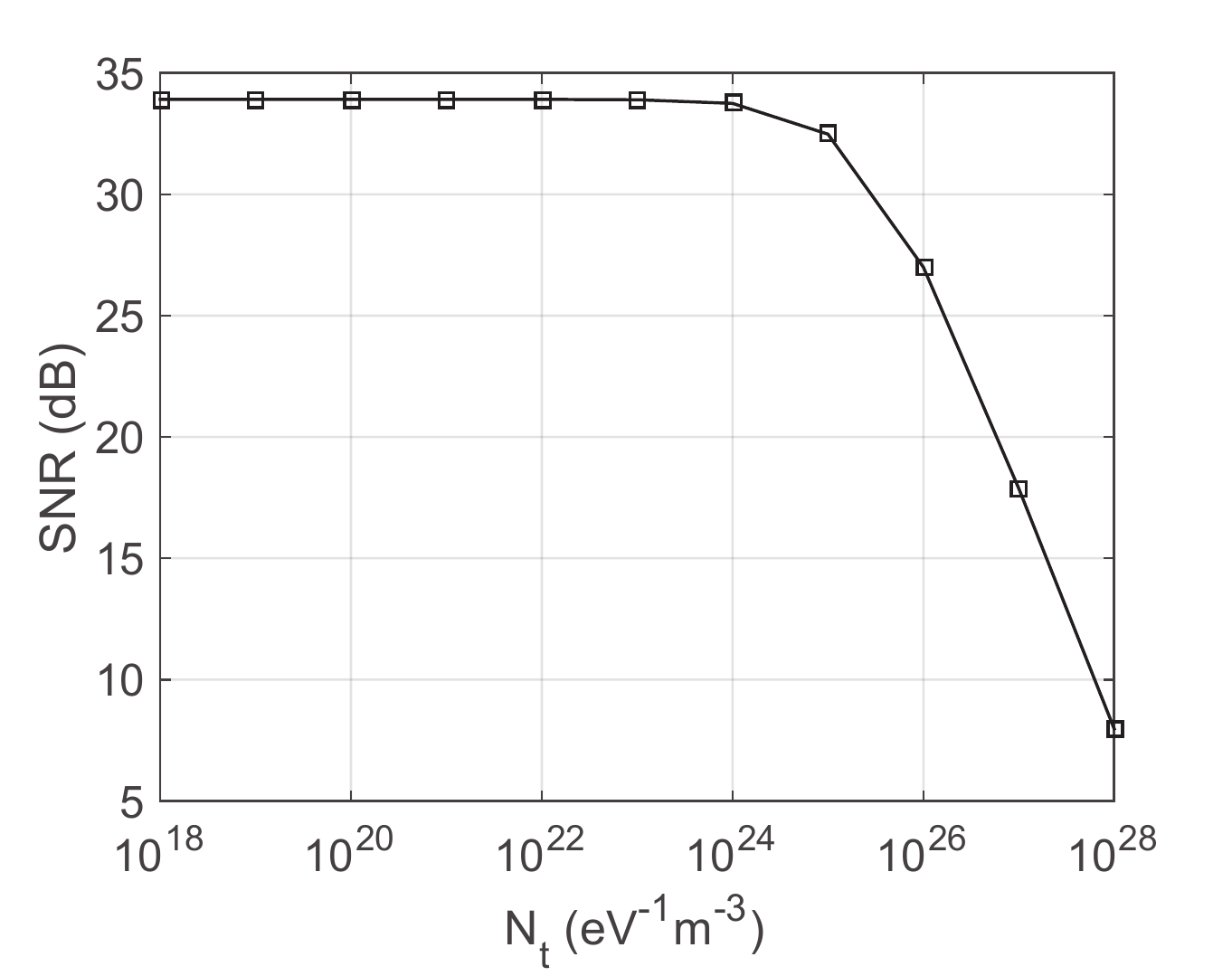}
 \label{fig:SNRNt}
 }
 \caption{SNR as a function of (a) ligand concentration $c_i$, (b) ionic concentration $c_{ion}$, (c) receptor length $L_R$, and (d) trap density $N_t$.}
 \label{fig:SNR}
 \end{figure*}

\subsubsection{SNR Analysis}
The receptor noise, which is the fluctuations in the number of bound receptors at steady-state, is reflected to the threshold voltage fluctuations by the relation $S_{\Delta V_{TH}^B}(f) = S_{\Delta N_B}(f) V_m^2$, where $V_m = (N_e q_{eff})/C_{eq,i}$ is the mean deviation in the threshold voltage resulting from the binding of a single ligand. Receptor and thermal noise can be assumed uncorrelated, since they are largely separated in frequency domain \cite{Spathis2015}. Including additive $1/f$ noise, overall PSD of the $\Delta V_{TH}$-referred noise can be given by \cite{Kuscu2015c}
\begin{equation}
S_{\Delta V_{TH}}(f) = S_{\Delta V_{TH}^B}(f) + S_{\Delta V_{TH}^T}(f) + S_{\Delta V_{TH}^F}(f).
\end{equation}
Additive noise effective on $\Delta V_{TH}$ is reflected to channel current noise via $S_{\Delta I_{DS}}(f) = S_{\Delta V_G}(f) g_m^2$. Assuming a resistance of 1$\Omega$ for the channel, we formulate the receiver SNR as follows
\begin{equation}
SNR = \frac{I_{DS}^2}{\int_{-\infty}^{\infty} S_{\Delta I_{DS}}(f) df}. \label{eq:SNRI}
\end{equation}
Using \eqref{eq:SNRI}, we investigate the effect of different system parameters on the SNR of the receiver's electrical output. SNR for varying ligand concentration corresponding to different symbols is plotted in Fig. \ref{fig:SNRc}, which clearly shows that SNR is significantly improved with increasing concentration. However, it begins to saturate at around $25$dB due to the saturation of the surface receptors for high ligand concentrations. The effect of ionic strength of the fluidic medium on the output SNR is given in Fig. \ref{fig:SNRcion}. When the ionic concentration increases above 100mol/m$^3$, the Debye length decreases below 1nm resulting in substantial screening of ligand charge. Therefore, SNR significantly decreases with increasing ionic strength. We also investigate the effect of receptor length on the SNR when the ionic strength is 70mM which makes the Debye length equal to 1.15nm. As seen in Fig. \ref{fig:SNRLr}, SNR in dB decreases linearly as the receptor length increases. Lastly, we analyze the SNR for varying trap density which is inversely proportional to the purity of the transducer channel. Trap density increases the $1/f$ noise, which is very effective in the frequency range of the antenna's operation. As is shown in Fig. \ref{fig:SNRNt}, the effect of trap density on the $1/f$ noise, and thus, on the SNR, is evident especially for $N_t > 10^{24}$eV$^{-1}$m$^{-3}$. Fortunately, experimentally reported trap densities for SiNW bioFETs are on the order of $10^{22}$eV$^{-1}$m$^{-3}$ \cite{Rajan2011a}.

\section{Advanced Design Issues}
\label{Advanced}

\subsection{Molecular Division Multiple Access (MDMA) and MoSK}
Use of different types of molecules brings up several opportunities to improve the performance of MC. MoSK, where messages are encoded into the type of molecules, is one of the two robust modulation techniques envisioned for MC \cite{Kuran2011}. Combining MoSK scheme with CSK, i.e., representing messages by both molecular concentration and type, multiplies the alphabet size, thus, can boost the communication rate. Moreover, employing various types of molecules for encoding can enable a particular code division multi-access scheme termed Molecular Division Multiple Access (MDMA) \cite{Gine2009}.

\begin{figure}[!t]
\centering
\includegraphics[width=7.5cm]{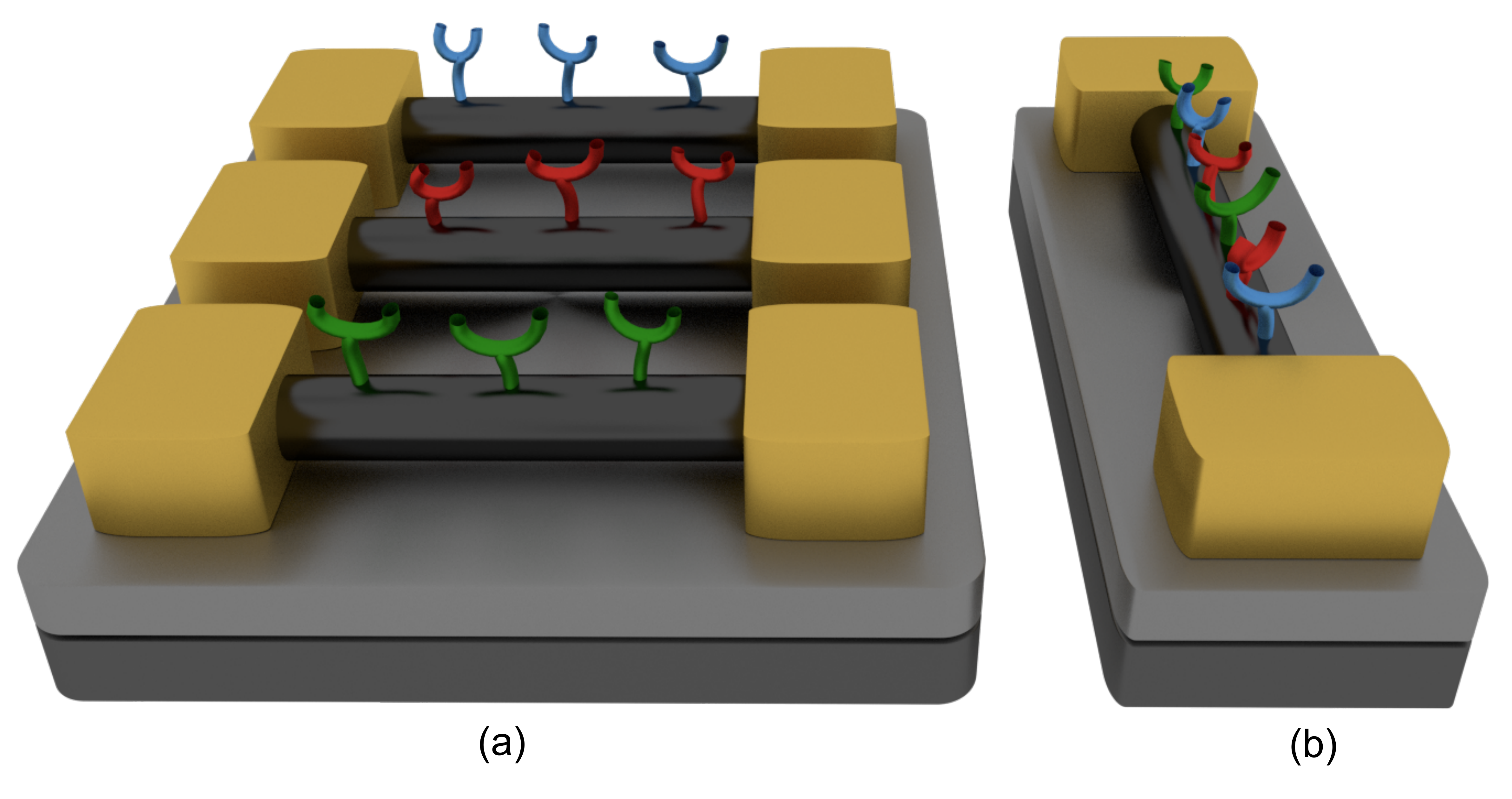}
\caption{Receiver antennas for MDMA and MoSK schemes.}
\label{fig:MDMA}
\end{figure}

All of these modulation and multiple access schemes require for different messages to be discriminated by the receiver. The receiver architecture based on bioFETs can be adapted to each of these schemes through one of the following modifications:
\begin{itemize}
\item Employing multiple antennas with different types of receptors each corresponding to a different ligand as shown in Fig.~\ref{fig:MDMA}(a), receiver can separately process the outputs of different antennas, and thus, easily decode different messages. Hence, it does not require a complex signal processing for detection; however, it implies a larger area of deployment, and more energy consumption which multiplies with the number of antennas.

\item Employing a single antenna is convenient if a significant diversity exists in efficient charges, $N_e q_{eff}$, among different ligands. The receptors can be of the same type if the ligands have similar affinity to the same receptor; otherwise, for each of the ligand type, different receptors should be deployed on the recognition layer as shown in Fig.~\ref{fig:MDMA}(b). It is clear that employing a single antenna is more energy-efficient and space-saving compared to the other architecture. However, for the case of MDMA where different signals received at the same time are superposed at the output, this architecture requires more complex signal processing on the electrical signal to discriminate the contributions of different messages \cite{Horesh2011}. 
\end{itemize}

\subsection{Diversity Combining}
The size of nanomachines could be significantly larger compared to a single receiver or transmitter antenna, and they are generally envisioned as motile undergoing both translational and rotational motion, which means that the distance between the transmitter and the receiver antennas of communicating nanomachines may alter notably during the operation. The distance between transmitter and receiver antennas has strong effect on the attenuation of molecular signals \cite{Akyildiz2013}. Hence, employing multiple receiver antennas with spatial diversity on a single nanomachine can lead to important diversity gains and reductions of the transmission delays. Several diversity combining schemes like selective combining, maximum ratio combining are applicable for the electrical receiver \cite{Meng2012}.

\subsection{Energy Efficiency Schemes}
Nanomachines are envisioned to have limited power; thus, energy consumption of the receiver should be minimized. This objective can be realized by optimizing the hardware, or employing energy-efficient scheduling algorithms.
Since the receiver is electrically powered, a nanomachine has the control on its operation times and sampling periods. Therefore, most of the conventional energy-efficient schemes in the literature can also be incorporated to the molecular receiver \cite{Akyildiz2002}. However, power saving modes of receiver's operation like sleep, idle, standby, and active modes, need to be defined in the context of the nanocommunications application.

\subsection{EM-Molecular Hybrid Operation}
Graphene and CNT, with extraordinary electromagnetic (EM) properties, have also proven to be promising candidates for the design of transceiver antennas in nanoscale EM communications \cite{Jornet2013a}. Designing the MC receiver with graphene or CNT-based transducer channels can allow a hybrid receiver architecture that can operate both in MC and EM nanocommunications. This hybrid design could enable novel applications. For example, the hybrid receiver can act as a nanoscale gateway that connects an \emph{in vivo} molecular nanonetwork to an external macroscale network, e.g., Internet, through an EM-based communication channel to enable Internet of Nano Things (IoNT) \cite{Akyildiz2015} \cite{Kuscu2015}. In a similar manner, two distant molecular nanonetwork can be connected with an EM link. However, additional challenges should be addressed. For example if the intention is to design a hybrid receiver antenna, the effect of biofunctionalization on the electromagnetic characteristics of the transducer channel should be analyzed.

\section{Challenges}
\label{Challenges}
\subsection{Fabrication Challenges}
Nanoscale semiconductors, which constitute the transducer channel, can be fabricated following any of the two fundamental approaches: bottop-up or top-down \cite{Lu2013}. Both approaches impose unique challenges to the proper fabrication of semiconductors. Although top-down methods are more costly compared to bottom-up synthesis, they provide more advanced control enabling mass production and compatibility with CMOS fabrication technologies promising for large-scale integration. However, top-down methods are known to be more amenable to producing defects and traps on channel surface, leading to higher level of $1/f$ noise \cite{Rajan2013-2}. On the other hand, bottom-up methods mostly rely on stochastic processes and do not allow an entire control over fabrication, leading to reproducibility problems, which is the main challenge to produce bioFETs with homogeneous characteristics \cite{Matsumoto2013}.

Immobilization of bioreceptors on the semiconductor channel can be realized through covalent or non-covalent binding, depending on the type of semiconductor and receptor molecules \cite{Lu2013}. Surface binding of receptors is a random process, and it is not possible with the current state-of-the-art technologies to exactly control the number of immobilized receptors and their orientation on the surface.
This implies another reproducibility problem that should be accounted for while implementing a bioFET-based receiver.

The effect of imperfect fabrication and the inevitable randomness in main system parameters, such as the concentration of surface receptors and the net charge of ligands, could be reflected as noise processes into the analytical model.

\subsection{Scaling and Integration Challenges}
Main functional units of the bioFET-based MC receiver, which are the biorecognition layer and the semiconductor channel, are already made of nanoscale materials; and thus, they do not pose a problem for scaling of the receiver. The major components that set the limitation for the receiver size are the source and drain contacts together with the reference electrode which is generally used for the aim of stabilizing the surface potential for improved detection performance \cite{Rajan2013-2}.

It is an ongoing challenge to miniaturize the reference electrode by keeping its stable operation. For the aim of integrating bioFETs to small scale on-chip devices, pseudo reference electrodes has been introduced with smaller sizes compared to a true reference electrode. However, they compromise voltage stability to some extent, and need to be further scaled down to work on a nanoscale device \cite{Rajan2013-2}. Recently, a promising approach has been proposed in \cite{Jain2012} to completely remove the need for a reference electrode. Combining the piezoelectric effect with the bioFET concept, the authors came up with a novel architecture named Flexure-FET which achieves to detect molecular concentration with high sensitivity, even by not requiring for ligands to be charged.

\subsection{Application Challenges}
A major challenge arises when the receiver is employed in an \textit{in vivo} application. The physiological conditions imply solutions with high ionic concentrations, abundance of interferers and contaminants, and existence of disruptive flows and fluctuating temperature, which may degrade the receiver's performance in several aspects. First, high ionic concentration, which is $\sim$100mM for physiological conditions, creates a strong screening effect reducing the Debye length to $\sim$1nm, and thus, impedes the sensitivity of the receiver \cite{Rajan2013}. Moreover, interferers that have affinities for the receptors could create strong background noise; and contaminants and disruptive flows may alter the binding kinetics, impede the stability of the receptors, even separate them from the dielectric layer.

Even though using highly charged ligands and very small size receptors like aptamers can solve the screening problem in theory, recently proposed frequency domain technique promises for much more realistic solutions. In \cite{Zheng2010}, the authors have shown that the power spectrum of binding events is Lorentzian-shaped, and easy to distinguish from the $1/f$ noise in frequency domain, because of the weakening of $1/f$ noise at high frequencies. They reveal that frequency domain detection outperforms the conventional time domain technique in terms of sensitivity in highly ionic solutions.

\section{Future Research Avenues}
\label{Future}
Progress in biosensor research has been mostly based on empirical studies, and the proposed designs generally lack the complementary analytical models. This leads to a problem for reproducibility of experimental results, and thus, to a plethora of studies that are not consistent with each other. There are only a few attempts to analytically model a biosensor, most of which are cited in this paper \cite{Hassibi2007} \cite{Rajan2013-2} \cite{Kuscu2015c}. However, these works together with our model assume that the ligands to be detected are in equilibrium with receptors. This is acceptable in the context of biosensors, because they are usually envisioned to be operated in laboratory conditions as being immersed in pre-prepared solutions with long incubation times. However, for the case of communications, considering the continuous and \textit{in situ} operation of the envisioned nanonetwork applications, this assumption could be violated when the bandwidth of the incoming signal is comparably high or the diffusion effects cannot be neglected. Therefore, the first research avenue should be development of comprehensive analytical models both for time and frequency domains that capture the transient dynamics, the effect of environmental conditions, also the mobility of nanomachines.

A limited number of works attempted to develop realistic simulation frameworks for diffusion-based MC and bioFETs. The current MC simulators are designed based on NS-3, and are mainly focused on the communication channel \cite{Llatser2014}. The models used in the simulators assume ideal MC receivers with perfect sampling of concentration. On the other hand, bioFET simulators are developed mainly for the purpose of proper evaluation and classification of experimental results \cite{Hediger2012}. These frameworks lack underlying analytical models, thus, are not able to provide the required design and optimization tools. At this point, a simulation framework for bioFET-based MC receivers needs to be developed using the analytical models; and then, the framework should be combined with the MC simulators, which are continuously being improved, to provide a more realistic simulation environment for end-to-end MC.

Analytical models and simulation frameworks should be accompanied by experimental validations. This requires the development of receiver prototypes based on bioFET architectures optimized for communication purposes. Microfluidics provides an invaluable platform to transport molecules in a controlled fashion. It has proven feasible for conducting MC experiments with controllable system parameters \cite{Bicen2013}, which makes microfluidics a promising means of further optimizing the MC receiver prototypes in real communication scenarios.

\section{Conclusion}
Physical design of a molecular receiver has been elaborated for the first time in the literature. Detection of messages in a molecular communication system mandates the label-free and \emph{in situ} sensing of molecules. A comprehensive review of the state-of-the-art biosensing methods revealed that the basic requirements of an MC receiver can be met only with a design in nanobioelectronic domain. FET-based biosensing mechanism is particularly favoured as the underlying mechanism of MC receiver due to the fast, reversible and simple operation. We conceptually presented the main operation principles of a FET-based MC receiver, and evaluated the reception performance in terms of sensitivity and SNR. These preliminary results demonstrating the reliable reception of molecular messages are encouraging for the further development of nanobioelectronic MC receiver. Future research avenues include the derivation of an improved analytical model for the receiver operation comprising the transient kinetics of molecular binding and transportation dynamics, optimization of the receiver performance for the envisioned applications through simulations and experiments, development of optimal detection schemes, and the integration of the receiver into a nanomachine.

\end{document}